\documentclass[aip,
 mph,%
 amsmath,
 amssymb,
 superscriptaddress,
 floatfix,
 preprint,
 eqsecnum,%
 numerical,%
]{revtex4-1}

\usepackage{graphicx}
\DeclareGraphicsExtensions{.eps,.png,.jpg}
\usepackage[caption=false%,labelformat=parens,subrefformat=subparens,listofformat=subparens
]{subfig}
\usepackage{epsfig}
\usepackage{epstopdf}
\usepackage{lipsum}
\usepackage{wrapfig}
\usepackage{wasysym}
\usepackage{threeparttable,booktabs,siunitx}
\usepackage{booktabs}
\usepackage{lmodern}
\usepackage{longtable}
\usepackage{multirow}
\usepackage{amsmath}
\usepackage[OT2,T1]{fontenc}
\usepackage[bookmarks=false]{hyperref}
\usepackage{color}
\hypersetup{pdfauthor={},pdfborder=0 0 0,colorlinks=true,citecolor=blue,linkcolor=blue,urlcolor=blue,}

\DeclareSymbolFont{cyrletters}{OT2}{wncyr}{m}{n}
\DeclareMathSymbol{\comb}{\mathalpha}{cyrletters}{"58}

\newcommand{\ben}{\begin{eqnarray}\displaystyle}
\newcommand{\een}{\end{eqnarray}}

\begin{document}

\title{Super resolution dual-energy cone-beam CT imaging with dual-layer flat-panel detector}
% Force line breaks with \\

\author{Ting Su}
\affiliation{Research Center for Medical Artificial Intelligence, Shenzhen Institutes of Advanced Technology, Chinese Academy of Sciences, Shenzhen, Guangdong 518055, China.}
\author{Jiongtao Zhu}
\affiliation{College of Physics and Optoelectronic Engineering, Ministry of Education and Guangdong Province, Key Laboratory of Optoelectronic Equipment and Systems, Shenzhen University, Shenzhen 518060, China.}
\author{Xin Zhang}
\affiliation{Research Center for Medical Artificial Intelligence, Shenzhen Institutes of Advanced Technology, Chinese Academy of Sciences, Shenzhen, Guangdong 518055, China.}
\author{Dong Zeng}
\affiliation{School of Biomedical Engineering, Southern Medical University, Guangzhou 510515, China.}%
\author{Yuhang Tan}
\affiliation{Research Center for Medical Artificial Intelligence, Shenzhen Institutes of Advanced Technology, Chinese Academy of Sciences, Shenzhen, Guangdong 518055, China.}
\author{Han Cui}
\affiliation{Research Center for Medical Artificial Intelligence, Shenzhen Institutes of Advanced Technology, Chinese Academy of Sciences, Shenzhen, Guangdong 518055, China.}
\author{Hairong Zheng}%
\affiliation{Paul C Lauterbur Research Center for Biomedical Imaging, Shenzhen Institutes of Advanced Technology, Chinese Academy of Sciences, Shenzhen, Guangdong 518055, China.}%
\author{Jianhua Ma}
\affiliation{School of Biomedical Engineering, Southern Medical University, Guangzhou 510515, China.}%
\author{Dong Liang}%
\affiliation{Research Center for Medical Artificial Intelligence, Shenzhen Institutes of Advanced Technology, Chinese Academy of Sciences, Shenzhen, Guangdong 518055, China.}
\affiliation{Paul C Lauterbur Research Center for Biomedical Imaging, Shenzhen Institutes of Advanced Technology, Chinese Academy of Sciences, Shenzhen, Guangdong 518055, China.}%
\author{Yongshuai Ge}%
\email{ys.ge@siat.ac.cn.}
\affiliation{Research Center for Medical Artificial Intelligence, Shenzhen Institutes of Advanced Technology, Chinese Academy of Sciences, Shenzhen, Guangdong 518055, China.}
\affiliation{Paul C Lauterbur Research Center for Biomedical Imaging, Shenzhen Institutes of Advanced Technology, Chinese Academy of Sciences, Shenzhen, Guangdong 518055, China.}%

\date{\today}% It is always \today, today,

\begin{abstract}
For medical cone-beam computed tomography (CBCT) imaging, the native receptor array of the flat-panel detector (FPD) is usually binned into a reduced matrix size. By doing so, the signal readout speed can be increased by over 4-9 times at the expense of sacrificing the spatial resolution by at least 50\%-67\%. Clearly, such tradition poses a main bottleneck in generating high spatial resolution and high temporal resolution CBCT images at the same time. In addition, the conventional FPD is also difficult in generating dual-energy CBCT images. In this paper, we propose an innovative super resolution dual-energy CBCT imaging method, named as suRi, based on dual-layer FPD (DL-FPD) to overcome these aforementioned difficulties at once. With suRi, specifically, an 1D or 2D sub-pixel (half pixel in this study) shifted binning is applied to replace the conventionally aligned binning to double the spatial sampling rate during the dual-energy data acquisition. As a result, the suRi approach provides a new strategy to enable high signal readout speed and high spatial resolution CBCT imaging with FPD. Moreover, a penalized likelihood material decomposition algorithm is developed to directly reconstruct the high resolution bases from the dual-energy CBCT projections containing spatial sub-pixel shifts. Experiments based on the single-layer FPD and DL-FPD are performed with physical phantoms and biological specimen to validate this newly developed suRi method. The synthesized monochromatic CT imaging results demonstrate that suRi can significantly improve the spatial image resolution by 46.15\%. We believe the developed suRi method would be capable to greatly enhance the imaging performance of the DL-FPD based dual-energy CBCT systems in future.
\end{abstract}
\maketitle
\section{Introduction}
\label{sec:introduction}
Over the past two decades, X-ray flat-panel detector (FPD) made from a single layer of CsI:TI scintillator and amorphous silicon ($\alpha$-Si) based thin film transistor (TFT) back-plate has been widely used in medical imaging applications. Due to its small native pixel dimension, e.g., 0.07-0.2 mm, FPD shows superior performance in detecting the ultra-fine details of the anatomical structures such as the micro-calcification, bone marrow and contrast enhanced blood vessels in two dimensional digital radiography (DR) imaging. For three and four dimensional cone-beam computed tomography (CBCT) imaging, flat-panel detector (FPD) is also irreplaceable in applications such as oral imaging\cite{schulze2020cone}, image-guided radiation therapy\cite{Oldham2005ConebeamCTGR} and interventional therapy \cite{Wallace2008ThreedimensionalCC}. Despite of these advancements, however, the dilemma between the intrinsic pixel-size-defined high spatial resolution and the pixel-number-confined slow data acquisition speed in FPD impedes its further developments in advanced CBCT imaging tasks which require both high spatial and high temporal resolution simultaneously. Take the PaxScan 4030CB FPD (Varex, USA) as an example, the native pixel dimension is 0.194~mm, and the full receptor array consists of 2048 $\times$ 1536 detector elements. With the $1\times1$ binning mode (limiting resolution 2.58 lp/mm), this FPD can only acquire 7.5 frames per second (fps) at most\cite{WinNT,sheth2022technical}, and at least 40 seconds are needed to complete a CBCT scan. The excessive CBCT scanning time would cause the motion artifacts and the loss of temporal variations of the contrast agent. Therefore, the FPD is usually worked at $2\times2$ binning mode (0.388~mm effective pixel dimension, corresponding to a reduced limiting spatial resolution of 1.29~lp/mm). By doing so, a quick signal readout speed of 30.0 fps can be achieved to complete the CBCT data acquisition within 6-10 seconds. Compared to the $1\times1$ binning mode, the $2\times2$ binning mode saves about 80\% CBCT scan time, but at the expense of losing 50\% image spatial resolution\cite{WinNT}. Besides, the current FPD is also difficult in performing dual-energy CBCT imaging to generate quantitative material-specific images for accurate disease diagnoses\cite{zbijewski2014dual}.

As a promising dual-energy CBCT imaging approach, the flat-panel detector made from dual layers of CsI:TI scintillator and $\alpha$-Si TFT back-plate (DL-FPD) can acquire the dual-energy CBCT data simultaneously: the low-energy (LE) X-ray projection is acquired from the top layer, and the high-energy (HE) X-ray projection is acquired from the bottom layer. The two CsI:TI scintillator layers may have different thicknesses, and additional beam filtration made of 1.0 mm Copper may also be inserted between the two layers to further separate the beam spectra. By far, several investigations based on the DL-FPD have been reported. 
For example, Shi \textit{et al.} and St{\aa}hl \textit{et al.} demonstrated the feasibility of performing accurate dual-energy CBCT imaging based on the DL-FPD \cite{shi2020characterization,staahl2021performance}. Wang \textit{et al.} proposed a model-based high resolution material decomposition method for the DL-FPD CBCT\cite{Wang2021HighresolutionMM}.  
Even though the DL-FPD helps realizing the dual-energy CBCT imaging, however, the conflict between the high spatial resolution imaging and the high speed imaging in single-layer FPD (SL-FPD) is still inherited. In other words, the DL-FPD encounters the same trade-off between the spatial resolution and the temporal resolution as the SL-FPD.

To overcome this long-standing difficulty, we propose an innovative imaging method, named as suRi, for DL-FPD to realize high temporal and spatial resolution CBCT imaging at the same time. In particular, the conventionally aligned binning approach is altered into a sub-pixel (half pixel in this study) shifted binning approach, see the illustrations in Fig.~\ref{fig1} for more details. Depending on the imaging task, such half pixel shift could be one-dimensional along the horizontal direction, or two-dimensional along the diagonal direction. Take the simple one-dimensional half pixel shift as an example, the information recorded on the top and bottom layers are spatially shifted by half-pixel. As shown in Fig.~\ref{fig1}, the line integral of the polychromatic X-ray beam that passes through a certain object position is sampled by two different FPD pixel elements: the top one is shifted by half pixel with regard to the bottom one. Consequently, the sub-pixel shift used by suRi doubles the spatial sampling rate assuming parallel incident beam. Therefore, the proposed suRi approach not only enables fast dual-energy CBCT imaging, but also enables super spatial resolution CBCT imaging. With such redefined DL-FPD data acquisition scheme, we believe the dual-energy CBCT imaging with high spatial resolution and high temporal resolution performance can be achieved. To the best of our knowledge, no such investigations have been reported before.

The major contributions of this work are as follows: (1) The sub-pixel shift dual-energy CBCT data acquisition scheme is proposed for DL-FPD for the first time. This method perfectly fits with the stacked structure of the DL-FPD to increase the spatial sampling rate while shortening the signal readout time by introducing additional ``geometric mismatch'' between the two detector layers. (2) A high performance one-step material decomposition algorithm is developed for the proposed novel DL-FPD based sub-pixel shift dual-energy CBCT imaging. The transmitted X-ray intensity of the sub-ray in each sub-pixel together with the CBCT imaging geometry, the data noise statistics and the beam spectra are utilized to build a more accurate forward imaging model, and the penalized likelihood function is optimized to generate the basis images with high spatial resolution. 

The rest of this paper is organized below: Section \ref{sec: related_work} introduces some related works. Section \ref{sec: method} presents the mathematical model of the sub-pixel shift data acquisition method for DL-FPD, the penalized likelihood material decomposition method, the experimental setup, the implementation details and evaluation metrics. Section \ref{sec:results} presents the experimental results of different objects. Section \ref{sec:conclusion} provides the discussions and a brief conclusion. 

\section{Related work} \label{sec: related_work}
\subsection{Super resolution CT imaging method}

There are two main strategies to realize super spatial resolution imaging in medical CBCT applications. First, the flying focal spot (FFS) technique \cite{Kachelriess2006FlyingFS,Flohr2005ImageRA} used by some advanced X-ray tube doubles the spatial sampling density in the horizontal and vertical directions, and thus can significantly reduce the in-plane and the out-of-plane aliasing artifacts. Since the detector matrix size is fixed, the FFS technique usually augments the total data size and adds burdens to the detector read-out speed. Second, the sub-pixel shifting technique in the detector end, which has been proposed for optical imaging for a long time \cite{Ur1992ImprovedRF}, and can also be utilized to greatly improve the spatial resolution of CT images. For example, Yan \textit{et al.} proposed to shift the single-layer detector array and make multiple scans to jointly reconstruct the high spatial resolution CT image \cite{Yan2015SuperRI}.  Li \textit{et al.} suggested to move the rotation stage in a fixed trajectory during the data acquisition to obtain multiple projection images with sub-pixel displacements \cite{Li2020MicroCTIO}. Szczykutowicz \textit{et al.} developed a spectral-spatial encoding method that acquires multiple measurements at different beam energies and different spatial positions to decompose the line integrals into two unique bases \cite{Szczykutowicz2021SubPR}. However, one major limitation of these studies is the prolonged data acquisition period due to the repeated measurements. In this work, the proposed suRi method can efficiently overcome such limitation by incorporating the sub-pixel shifts into the DL-FPD.

\subsection{Material decomposition algorithm}
The dual-energy CT material decomposition methods can be mainly divided into three categories. The first one is known as the projection-domain material decomposition method\cite{Alvarez1976EnergyselectiveRI,Schlomka2008ExperimentalFO}, which discriminates the material-specific density integrals from paired dual-energy projections. This kind of method is difficult to be implemented straightforwardly in this study due to the spatial mismatch between the dual-energy projections. The second one is the image-domain decomposition method \cite{Niu2014IterativeID,Mendona2014AFM,Zeng2020FullSpectrumKnowledgeAwareTM}, in which the CT images reconstructed using the filtered back-projection (FBP) or iterative algorithm are decomposed into different bases. Despite that the spatial mismatch of projections can be compensated after reconstruction, however, the image-domain method is not able to generate CT bases with the desired high spatial resolution. The third approach is the so called direct (one-step) material decomposition method\cite{Mory2018ComparisonOF,Chen2018AlgorithmenabledPC,6805661}, which generates the material-specific CT maps directly from the dual-energy projections by using a certain imaging model. Since the the imaging geometry (e.g., the sub-pixel shifts), the noise fluctuation and the beam spectra can be incorporated into the formulated imaging model, therefore, the direct material decomposition method is selected in this study to reconstruct high spatial resolution CT bases directly from the acquired sub-pixel shifted dual-energy projections.

\begin{figure*}[htb]
\centerline{\includegraphics[width=18cm]{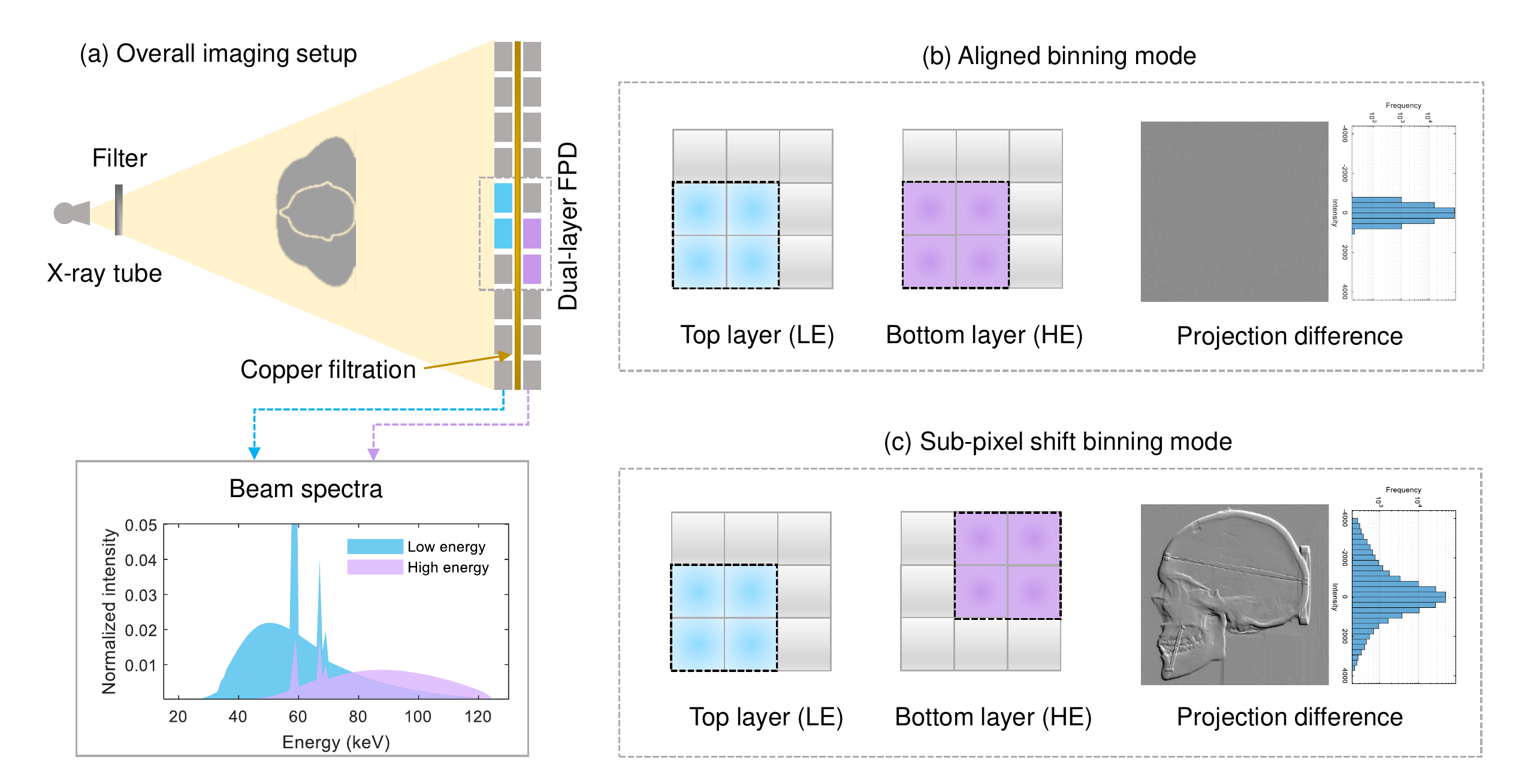}}
\caption{Illustration of the proposed suRi method in this study. (a) The overall imaging setup of the DL-FPD based DECT system. In DL-FPD, the top layer and bottom layer record the low-energy and high-energy X-ray photons, respectively. (b) The conventionally used data acquisition scheme, denoted as aligned binning, in which the signal of four neighboring pixels  at the same spatial locations of the top and bottom detector layers are summed up. (c) The proposed suRi data acquisition scheme, in which the binned pixels in the top layer and bottom layer use are shifted diagonally by half pixel size. As clearly seen from the difference image, inconsistent spatial information are recorded in the top layer and bottom layer, separately. 
Due to this reason, the proposed suRi method can be used to restore the high spatial resolution information and thus achieve high spatial resolution DECT imaging. Note that 2 $\times$ 2 binning is assumed in (b) and (c) to demonstrate the aligned binning and sub-pixel shift binning modes.}
\label{fig1}
\end{figure*}

\section{Method} \label{sec: method}
\subsection{Imaging model in suRi}
\begin{figure}[htb]
\centerline{\includegraphics[width=8cm]{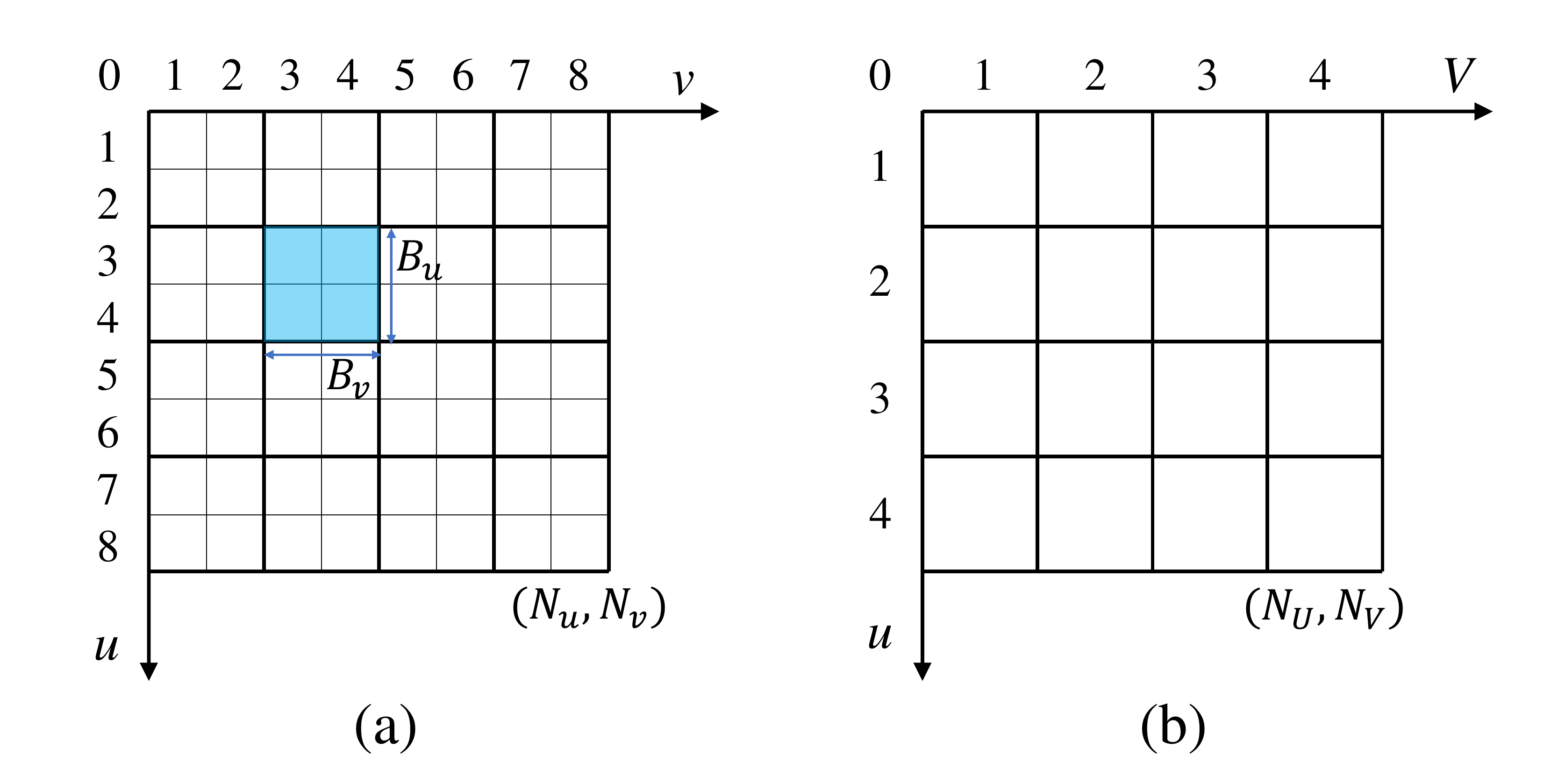}}
\caption{The discrete coordinates for (a) the high resolution unbinned projection data $\hat y_{u,v}^{'k}$, and (b) the low resolution binned projection data $\hat y_{U,V}^{k}$. The binning mode $B_u\times B_v$ is applied to generate $\hat y_{U,V}^{k}$ from $\hat y_{u,v}^{'k}$. Quantitatively, $N_u=B_u\times N_U$, $N_v=B_v\times N_V$. }
\label{fig02}
\end{figure}

To better explain the proposed suRi data acquisition method, we need to define several coordinate systems on the detector planes, see Fig.~\ref{fig02}. First, the coordinates of the unbinned detector elements are denoted by $(u,v)$, and they are assumed to be spatially fixed for all of the $K$ $(K\ge 2)$ detector layers, see Fig.~\ref{fig02}(a). Second, the coordinates of the $B_u\times B_v$ binned detector elements are denoted by $(U,V)$, see Fig.~\ref{fig02}(b). As a consequence, the beam intensity recorded by the binned detector element $(U,V)$ in the $k$th detector layer, denoted as $\hat y_{U,V}^k$, can be expressed as a summation of the responses $\hat y_{u,v}^{'k}$ that are acquired from the unbinned detector elements, namely,
	\begin{equation}
	\hat y_{U,V}^k = \sum_{u,v \in \mathcal{N}_{U,V}^k} \hat y_{u,v}^{'k},
	\label{eq01}
	\end{equation}
where $\mathcal{N}_{U,V}^k$ represents the neighboring $B_u\times B_v$ detector elements that belong to the binned detector element $(U,V)$ on the $k$th detector layer.

To determine $\mathcal{N}_{U,V}^k$, we assume that the binned detector element $(U,V)$ is spatially displaced between two adjacent detector layers:
\begin{equation}
u_{i}^{(k+1)}(U,V) - u_{u}^{(k)}(U,V)= \Delta_u^k \times \delta_{del},
\label{eq02}
\end{equation}
\begin{equation}
u_{v}^{(k+1)}(U,V) - u_{v}^{(k)}(U,V)= \Delta_v^k \times \delta_{del},
\label{eq02}
\end{equation}
where $u_{u}^{(k)}(U,V)$ and $u_{v}^{(k)}(U,V)$ are the vertical and horizontal physical locations of the binned detector element $(U,V)$ in the $k$th layer, $\delta_{del}$ denotes the physical dimension of one native detector element, $\Delta_u^k$ and $\Delta_v^k$ denote the relative displacements along the vertical and horizontal directions, respectively. Without loss of generality, herein we assume $\Delta_u^k\ge0$ and $\Delta_v^k\ge0$. It is easy to derive that the coordinates of the neighboring pixels in $\mathcal{N}_{U,V}^k$ satisfies:
	\begin{equation}
	(U-1)\times B_u+1+\sum_{\kappa=1}^k\Delta_u^\kappa\leq u \leq U\times B_u+\sum_{\kappa=1}^k\Delta_u^\kappa 
	\end{equation}
	\begin{equation}
	(V-1)\times B_v+1+\sum_{\kappa=1}^k\Delta_v^\kappa\leq v \leq V\times B_v+\sum_{\kappa=1}^k\Delta_v^\kappa
	\end{equation}
The above two relationships define the spatial correspondence between the native pixel $(u, v)$ and the corresponding binned element $(U,V)$. In particular, if $\Delta_u^k=\Delta_v^k=0$, the equality $\mathcal{N}_{U,V}^1 = ...=\mathcal{N}_{U,V}^K$ holds for the K number of detector layers. Under such circumstance, the imaging model provided in (\ref{eq01}) corresponds to the aligned binning data acquisition, as schemed in Fig.~\ref{fig1}(b). However, if the displacements satisfies $0<\Delta_u^{k\ge2}<B_u$ or $0<\Delta_v^{k\ge2}<B_v$ (herein, we assume $\Delta_u^{k=1}=\Delta_v^{k=1}=0$, meaning the element binning on the first detector layer is not displaced.), then $\mathcal{N}_{U,V}^k$ varies from $\mathcal{N}_{U,V}^{k+1}$, indicating that some detector elements are overlapped, as schemed in Fig.~\ref{fig1}(c). Mathematically, the indices of $u,v \in \mathcal{N}_{u,v}^k\cap\mathcal{N}_{U,V}^{k+1}$, are found to be:
	\begin{equation}
	(U-1)\times B_u+1+\sum_{\kappa=1}^{k+1}\Delta_u^\kappa\leq u \leq U\times B_u+\sum_{\kappa=1}^k\Delta_u^\kappa 
	\end{equation}
	\begin{equation}
	(V-1)\times B_v+1+\sum_{\kappa=1}^{k+1}\Delta_v^\kappa\leq v \leq V\times B_v+\sum_{\kappa=1}^k\Delta_v^\kappa
	\end{equation}
The transition from the unbinned projection $\boldsymbol {\hat Y'}^{k}\in \mathbb{R}^{N_uN_v}$ of the $k$th detector layer,
\begin{equation}
	\boldsymbol {\hat Y'}^k=[\hat y_{1,1}^{'k},\hat y_{2,1}^{'k},...,\hat y_{1,2}^{'k},\hat y_{2,2}^{'k},...,\hat y_{u,v}^{'k},...\hat y_{N_u-1,N_v}^{'k},\hat y_{N_u,N_v}^{'k} ]^T
	\end{equation}
to the binned projection $\boldsymbol {\hat Y}^k\in \mathbb{R}^{N_UN_V}$
	\begin{equation}
	\boldsymbol {\hat Y}^k=[\hat y_{1,1}^k,\hat y_{2,1}^k,...,\hat y_{1,2}^k,\hat y_{2,2}^k,...,\hat y_{U,V}^k,...\hat y_{N_U-1,N_V}^k,\hat y_{N_U,N_V}^k ]^T
	\label{eq05}
	\end{equation}
can be derived based upon (\ref{eq01}):
	\begin{equation}
	\boldsymbol {\hat Y}^k=\boldsymbol D^k\boldsymbol {\hat Y}^{'k},
	\label{eq06}
	\end{equation}		
where $\boldsymbol  D^k\in \mathbb{R}^{N_UN_V\times N_uN_v}$ denotes the displaced binning matrix and is composed by 0 and 1. 

A simple example is discussed below to demonstrate that the unbinned projection $\boldsymbol {\hat Y'}^{k}\in \mathbb{R}^{N_uN_v}$ may be recovered from the binned projection $\boldsymbol {\hat Y}^k\in \mathbb{R}^{N_UN_V}$. We assume a linear dual-layer (K=2) detector array with 8 native elements ($N_u=1$, $N_v=8$) in each layer, and the $1\times2$ binning ($B_u=1$, $B_v=2$) is applied. Thus, the binned detector matrix has a size of $1\times4$ ($N_U=1$, $N_V=4$) for each detector layer. To avoid the illness of the matrix inversion in (\ref{eq06}), the displacement of binned detector element is set to $\Delta_u^k=0$ and $\Delta_v^k=1$. Hence, the displacement matrix $\boldsymbol  D^k$ in each detector layer is written as:
	\begin{eqnarray}
	\boldsymbol  D^1=
	\begin{bmatrix}
	1&1&0&0&0&0&0&0\\
	0&0&1&1&0&0&0&0\\
	0&0&0&0&1&1&0&0\\
	0&0&0&0&0&0&1&1\\
	\end{bmatrix},\\
	\boldsymbol  D^2=
	\begin{bmatrix}
	0&1&1&0&0&0&0&0\\
	0&0&0&1&1&0&0&0\\
	0&0&0&0&0&1&1&0\\
	0&0&0&0&0&0&0&1\\
	\end{bmatrix}. 
	\label{eq07}
	\end{eqnarray}
Ignoring the difference of X-ray beam spectra recorded by the two detector layers, i.e., $\boldsymbol {\hat Y}^{'1}=\boldsymbol {\hat Y}^{'2}=\boldsymbol {\hat Y}^{'}$ for the monochromatic X-ray beam, consequently, we obtain:
	\begin{equation}
		\begin{bmatrix}
			\boldsymbol {\hat Y}^1\\
			\boldsymbol {\hat Y}^2\\
			\end{bmatrix}
		=	
		\begin{bmatrix}
			\boldsymbol  D^1\\
			\boldsymbol  D^2\\
		\end{bmatrix}
		\times
		\boldsymbol {\hat Y}^{'}.
		\label{eq08}	
	\end{equation}
With a standard matrix inversion, the unbinned $\boldsymbol {\hat Y}^{'}$ corresponding to the high resolution projection can be readily obtained from the two binned projections $\boldsymbol {\hat Y}=[\boldsymbol {\hat Y}^1, \boldsymbol {\hat Y}^2]^T$:
	\begin{equation}
	\boldsymbol {\hat Y}^{'}=\boldsymbol  D^{-1}\boldsymbol {\hat Y}
	\label{eq09}
	\end{equation}
%Obviously, (\ref{eq08}) can also be extended into two-dimensional area multi-layer FPDs and detector element displacements. \textcolor{red}{Notice that the $\Delta_u^{k\ge2}=1$ and $\Delta_v^{k\ge2}=1$ have to be assumed if the native unbinned $\boldsymbol {\hat Y}^{'}$ is recovered. In other words, the spatial displacements between the two binned detector elements having the identical index $(U,V)$ on the adjacent layers is set equal to $\delta_{del}$ along both directions.}
%
%\textcolor{red}{Such above requirement is easy to be implemented by the newly proposed sub-pixel shifted data acquisition method suRi with multi-layer FPD in this work. For example, projections having $\delta_{del}\times\delta_{del}$ detector element size can be recovered from the $2\times2$ binned DL-FPD with $\delta_{del}\times\delta_{del}$ adjacent spatial displacements, and projections having $\delta_{del}\times\delta_{del}$ detector element size can be recovered from the $3\times3$ binned triple-layer FPD with $\delta_{del}\times\delta_{del}$ adjacent spatial displacements. More examples can be found in Table~\ref{table0}. Additionally, the potential acceleration rate of signal readout speed is also estimated.}

With the sub-pixel shift technique, the spatial resolution of the projection image obtained from the DL-FPD or ML-FPD can be potentially improved by a factor of $K$ (along the $u$ or $v$ direction). Table~\ref{table0} provides more examples using the dual-layer and tripe-layer FPD. Additionally, the potential acceleration rate of signal readout speed is also estimated. Be aware that only the 1D horizontal sub-pixel shifts are considered.

\begin{table}[htb]
\caption{Detector settings in suRi with 1D horizontal sub-pixel shift. The physical shift and resolution are listed along the $u$ and $v$ directions. The potential acceleration rate of signal readout speed is compared with the $1\times1$ binning mode.}
\label{table0}
\begin{tabular}{ccccc}
\hline\hline
Detector                      & Binning    & Physical shift           & Resolution                            & Acceleration \\ \hline
\multirow{3}{*}{Dual-layer}   & 1$\times$2 & {[}0, $\delta_{del}${]}  & {[}$\delta_{del}$, $\delta_{del}${]}  & 2            \\
                              & 1$\times$4 & {[}0, $2\delta_{del}${]} & {[}$\delta_{del}$, $2\delta_{del}${]} & 4            \\
                              & 1$\times$6 & {[}0, $3\delta_{del}${]} & {[}$\delta_{del}$, $3\delta_{del}${]} & 6            \\ 
\multirow{3}{*}{Triple-layer} & 1$\times$3 & {[}0, $\delta_{del}${]}  & {[}$\delta_{del}$, $\delta_{del}${]}  & 3            \\
                              & 1$\times$6 & {[}0, $2\delta_{del}${]} & {[}$\delta_{del}$, $2\delta_{del}${]} & 6            \\
                              & 1$\times$9 & {[}0, $3\delta_{del}${]} & {[}$\delta_{del}$, $3\delta_{del}${]} & 9            \\ \hline\hline
\end{tabular}
\end{table}

In reality, it is difficult to solve Eq.~(\ref{eq06}) due to the different beam spectra recorded by the individual detector layer in DL-FPD. Nevertheless, it is still possible to improve the CT image spatial resolution via suRi since the needed high spatial resolution information have already been encoded in the acquired projections. To reconstruct such CT images with high spatial resolution, the iterative CT reconstruction algorithm has to be employed.

\subsection{Penalized likelihood material decomposition}	\label{Subsec:algorithmn}

In CT imaging, the expected intensity of the $i$th ray registered by a binned detector pixel is treated as a sum of those obtained from several discrete sub-rays $r$ (corresponding to different sub-pixels). The linear attenuation coefficient of a compound material can be regarded as the weighted combination of different basis materials. Thus (\ref{eq01}) can be rewritten as:
	\begin{equation}
	\hat y_{i}^k (\boldsymbol \rho)= \sum_{r \in \mathcal{N}_i^k} \sum_{e = 1}^E S_{ie}^{kr} {\rm exp} \left(-\sum_{m = 1}^M f_e^m \sum_{j=1}^J a_{ij}^{kr} \rho_j^m\right).
	\label{eq4}
	\end{equation}
where $S_{ie}^{kr}$ denotes the spectral intensity at energy $e$ for the $r$th sub-ray of ray $i$, $f_e^m$ denotes the mass attenuation coefficient of material $m$ at energy $e$, $a_{ij}^{kr}$ denotes the intersection length of the $r$th sub-ray of ray $i$ with the $j$th voxel, $\rho_j^m$ denotes the density value of material $m$ at voxel $j$, $\boldsymbol \rho=(\rho_1^1,..., \rho_J^1;...; \rho_1^m,..., \rho_J^m;...;\rho_1^M,...,\rho_J^M)^T$ is a matrix of the density values. 
Assuming that the projection measurements follow Poisson distributions, we can obtain the negative log-likelihood function to solve for the unknown material basis maps: 
	\begin{equation}
	-L(\boldsymbol \rho) = -{\rm log}\left(\mathcal{P}(y | \hat y)\right)= \sum_{k = 1}^K \sum_{i=1}^I \hat y_{i}^{k} - {y_{i}^{k}} {\rm ln} \hat y_{i}^{k}.
	\label{eq6}
	\end{equation}
where $y_i^k$ denotes the measurement of the $i$th ray from layer $k$. To suppress the image noise, a regularization term $R(\boldsymbol \rho)$ is added, resulting in the following penalized likelihood objective function: 
	\begin{equation}
	\Phi(\boldsymbol \rho) = -L(\boldsymbol \rho) + R(\boldsymbol \rho).
	\label{eq7}
	\end{equation}
Herein, the edge-preserving Huber regularization is utilized:	
	\begin{equation}
	R(\boldsymbol \rho) = \sum_{m = 1}^M \sum_{j = 1}^J \sum_{j' \in \mathcal{V}_j} \lambda^m \psi(\rho_j^m-\rho_{j'}^m),
	\label{eq8}
	\end{equation}
with		
	\begin{equation}
	\psi(\Delta) = \left\{
	             \begin{array} {lr}	
			\frac{\Delta^2}{2} \qquad \qquad \quad  {\rm if}\ |\Delta| \leq \gamma \\
			\gamma |\Delta|-\frac{\gamma^2}{2} \qquad {\rm if}\ |\Delta| > \gamma
	             \end{array}
	             \right.,
	\label{eq9}
	\end{equation}
in which $\lambda^m$ is a factor that balances the regularization weight between different basis materials, $\mathcal{V}_j$ is a neighborhood of the voxel $j$, $\gamma$ is a tuning parameter that decides whether the regions use quadratic regularization (when the neighborhood difference is smaller than $\gamma$) or linear regularization (when the neighborhood difference is larger than $\gamma$).

%\subsection{Optimization algorithm}
Since the objective function in (\ref{eq7}) is difficult to minimize directly, therefore, the optimization transfer principles\cite{Lange2000OptimizationTU,Pierro1995AME} is used to optimize a surrogate function, which is much more easier and efficient to minimize. The derivations of the surrogate functions are based on \cite{Mechlem2018JointSI} and \cite{Weidinger2016PolychromaticIS}. In addition, an extra surrogate function that extracts the sub-ray related summation outside the negative log-likelihood objective function is introduced as well. To simplify the derivation, we define
	\begin{eqnarray}	
	 h_i^k(\hat y_i^k) = \hat y_{i}^{k} - {y_{i}^{k}} {\rm ln} \hat y_{i}^{k},\quad \hat y_i^k = \sum_{r \in \mathcal{N}_i^k} \hat y_i^{'kr},\cr
	l_{ie}^{krm}(\boldsymbol \rho^m) = f_e^m \sum_{j=1}^J a_{ij}^{kr} \rho_j^m, \qquad \quad l_{ie}^{kr}(\boldsymbol \rho) = \sum_{m = 1}^M  l_{ie}^{krm},\cr
	t_{ie}^{kr}(\boldsymbol \rho) = {\rm exp} \left(-l_{ie}^{kr}\right),\cr
	\theta_i^{kr,(n)}=\frac{\hat y_i^{k,(n)}}{\hat y_i^{'kr,(n)}}, \quad \beta_{ie}^{kr,(n)} = \frac{\hat y_i^{'kr,(n)}}{t_{ie}^{kr,(n)}}.
	\label{eq11}
	\end{eqnarray}
Since it holds that: 
	\begin{equation}	
	\sum_{r \in \mathcal{N}_i^k} \frac{1}{\theta_i^{kr,(n)}} = 1, \quad \sum_{e=1}^E \frac{S_{ie}^{kr}}{\beta_{ie}^{kr,(n)}} = 1, 
	\label{eq12}
	\end{equation}
according to the Jensen's inequality \cite{mcshane1937jensen}, the integrals over the sub-ray $r$ and spectrum $k$ can be moved out of the negative log-likelihood function successively. As a consequence, two surrogate functions $Q_1(\boldsymbol{\rho}; \boldsymbol{\rho}^{(n)})$ and $Q_2(\boldsymbol{\rho}; \boldsymbol{\rho}^{(n)})$ can be obtained immediately:
	\begin{eqnarray}
	-L(\boldsymbol \rho) &=& \sum_{k=1}^K\sum_{i=1}^I h_i^k\left(\sum_{r \in \mathcal{N}_i^k} \hat y_i^{'kr}\right) \cr
	\leq \sum_{k=1}^K &\sum_{i=1}^I& \frac{1}{\theta_i^{kr,(n)}} \sum_{r \in \mathcal{N}_i^k} h_i^{kr}\left(\hat y_i^{'kr}\theta_i^{kr,(n)}\right)\equiv Q_1(\boldsymbol{\rho}; \boldsymbol{\rho}^{(n)}),  
	\label{eq13} 
	\end{eqnarray} 
and
	\begin{eqnarray}
	&Q_1&(\boldsymbol{\rho}; \boldsymbol{\rho}^{(n)}) = \sum_{k=1}^K\sum_{i=1}^I \frac{1}{\theta_i^{kr,(n)}} \sum_{r \in \mathcal{N}_i^k} h_i^{kr}\left(  \sum_{e=1}^E S_{ie}^{kr}t_{ie}^{kr}\theta_i^{kr,(n)} \right) \cr
	&\leq &\sum_{k=1}^K \sum_{i=1}^I  \sum_{r \in \mathcal{N}_i^k} \sum_{e=1}^E \frac{S_{ie}^{kr}}{\theta_i^{kr,(n)}\beta_{ie}^{kr,(n)}} h_{ie}^{kr}\left(   t_{ie}^{kr}\beta_{ie}^{kr,(n)}\theta_i^{kr,(n)} \right)  \cr
	&\equiv & Q_2(\boldsymbol{\rho}; \boldsymbol{\rho}^{(n)}).
	\label{eq14} 
	\end{eqnarray} 
Subsequently, $Q_2(\boldsymbol{\rho}; \boldsymbol{\rho}^{(n)})$ is approximated into a quadratic surrogate function in terms of the attenuation line integral $l_{ie}^{krm}$. With the definition of $g_{ie}^{kr}(l_{ie}^{krm}) \equiv h_{ie}^{kr}\left(  t_{ie}^{kr}\beta_{ie}^{kr,(n)}\theta_i^{kr,(n)} \right)$, $g_{ie}^{kr}$ can be expanded into a Taylor series about the line integral $l_{ie}^{krm}$ up to the second order:
	\begin{eqnarray}
	&g_{ie}^{kr}&(l_{ie}^{krm}) \approx q_{ie}^{kr}(l_{ie}^{krm}; l_{ie}^{krm,(n)}) = g_{ie}^{kr}(l_{ie}^{krm,(n)}) \cr
	&+& \sum_{m=1}^M \left.\frac{\partial g_{ie}^{kr} }{\partial l_{ie}^{krm}} \right|_{l_{ie}^{krm} = l_{ie}^{krm,(n)}}(l_{ie}^{krm} - l_{ie}^{krm,(n)}) \cr
	&+& \frac{1}{2} \sum_{c,d=1}^M T_{ie}^{krcd,(n)} (l_{ie}^{krc} - l_{ie}^{krc,(n)})(l_{ie}^{krd} - l_{ie}^{krd,(n)}),
	\label{eq15}
	\end{eqnarray}
where $l_{ie}^{krm,(n)} = l_{ie}^{krm}(\boldsymbol{\rho}^{m,(n)})$. Since it is difficult to choose a curvature $T_{ie}^{krcd,(n)}$ to fulfill all the conditions in the optimization transfer principles, therefore, the second derivative of $g_{ie}^{kr}(l_{ie}^{krm})$ is applied to replace the curvature by relaxing the monotonicity according to \cite{Erdogan1999OrderedSA,Weidinger2016PolychromaticIS}. Combining \eqref{eq14} and \eqref{eq15}, the third surrogate function $Q_3(\boldsymbol{\rho}; \boldsymbol{\rho}^{(n)})$ is obtained:
		\begin{equation}	
	Q_3(\boldsymbol{\rho}; \boldsymbol{\rho}^{(n)})=\sum_{k=1}^K \sum_{i=1}^I  \sum_{r \in \mathcal{N}_i^k} \sum_{e=1}^E \frac{S_{ie}^{kr}}{\theta_i^{kr,(n)}\beta_{ie}^{kr,(n)}}q_{ie}^{kr}(l_{ie}^{krm}; l_{ie}^{krm,(n)}).
	\label{eq16}
	\end{equation}
Next, the line integral $l_{ie}^{krm}$ is made separable over discrete density voxels $\rho_j^m$. According to \cite{Mechlem2018JointSI}, the surrogate function $Q_4(\boldsymbol{\rho}; \boldsymbol{\rho}^{(n)})$ can be obtained: 

	\begin{equation}	
	Q_4(\boldsymbol{\rho}; \boldsymbol{\rho}^{(n)})=\sum_{k=1}^K \sum_{i=1}^I  \sum_{r \in \mathcal{N}_i^k} \sum_{e=1}^E \sum_{j=1}^J \frac{\omega_{ij}^{kr}S_{ie}^{kr}}{\theta_i^{kr,(n)}\beta_{ie}^{kr,(n)}}q_{ie}^{kr}\left(\xi_{ije}^{krm}(\rho_j^m;\rho_j^{m,(n)})\right),
	\label{eq19}
	\end{equation}
where
	\begin{eqnarray}		
	\omega_{ij}^{kr}=\frac{a_{ij}^{kr}}{\sum_j^J a_{ij}^{kr}}, \cr
	\xi_{ije}^{krm}(\rho_j^m;\rho_j^{m,(n)})=\frac{a_{ij}^{kr}f^m}{\omega_{ij}^{kr}}\left(\rho_j^m-\rho_j^{m,(n)}\right) +l_{ie}^{krm,(n)}.
	\label{eq18}
	\end{eqnarray}
	A separable surrogate function for the regularization term can be derived \cite{Erdogan1999OrderedSA} as follows:
	\begin{eqnarray}
	\mathcal{R}(\boldsymbol \rho; \boldsymbol{\rho}^{(n)}) &=& \frac{1}{2}\sum_{m = 1}^M \sum_{j = 1}^J \sum_{j' \in \mathcal{V}_j} \lambda^m \left[ \psi(2\rho_j^m-\rho_j^{m,(n)}-\rho_{j'}^{m,(n)})\right. \cr
	&+&  \left. \psi(2\rho_{j'}^m-\rho_j^{m,(n)}-\rho_{j'}^{m,(n)}) \right].
	\label{eq20}
	\end{eqnarray}

With the above separable surrogate functions for both $-L(\boldsymbol \rho)$ and $R(\boldsymbol \rho)$ in \eqref{eq7}, the final surrogate objective function becomes $Q_4(\boldsymbol \rho; \boldsymbol{\rho}^{(n)})+\mathcal{R}(\boldsymbol \rho; \boldsymbol{\rho}^{(n)})$. We use the Newton-Raphson method to iteratively minimize the objective function:
	\begin{equation}	
	\boldsymbol\rho^{(n+1)} = \boldsymbol\rho^{(n)}-(H_{Q_4}^{(n)}+H_{\mathcal{R}}^{(n)})^{-1}\cdot\left.(\nabla Q_4+\nabla \mathcal{R})\right|_{\boldsymbol\rho=\boldsymbol\rho^{(n)}},
	\label{eq21}
	\end{equation}
where $H_{Q_4}^{(n)}$ and $H_{\mathcal{R}}^{(n)}$ denote the Hessian matrices of $Q_4(\boldsymbol \rho; \boldsymbol{\rho}^{(n)})$ and $\mathcal{R}(\boldsymbol \rho; \boldsymbol{\rho}^{(n)})$ at point $\boldsymbol \rho=\boldsymbol{\rho}^{(n)}$, $\nabla Q_4$ and $\nabla \mathcal{R}$ denote the gradient vectors. Detailed expressions can be found in the Appendix. 
To this end, the penalized likelihood objective function defined in (\ref{eq7}) can be minimized and optimized readily. To reduce the computation burden in each iteration and to accelerate the convergence speed, the ordered subset optimization \cite{Erdogan1999OrderedSA} and Nesterov’s momentum technique \cite{Mechlem2018JointSI} are applied.

%\begin{table}[htb]
%\caption{Pseudo-code of the proposed material decomposition algorithm.}
%\label{table1}
%\setlength{\tabcolsep}{3pt}
%\begin{tabular}{p{240pt}}
%\hline
%Initialization: $\rho_j^{m} = 0$, $H_{reg}^{juv}=0$ \\
%Precomputation: $A_{j}^{kr} = \sum_{i=1}^I  a_{ij}^{kr}$,  \\
%for each iteration $n=1:N$ \\
%\qquad for each subset $m=1:M$ \\
%%\qquad\qquad $A_m^{kr} = subset(A^{kr}, m)$\\
%%\qquad\qquad $y_m^{kr} = subset(y^{kr}, m)$\\
%\qquad\qquad $l_{ie}^{kr} = \sum_b f_e^m\sum_j a_{ij}^{kr} \rho_j^m$,  \qquad $\forall i \in Subset_m$\\
%\qquad\qquad $t_{ie}^{kr} = {\rm exp} \left(-l_{ie}^{kr}\right)$\\
%\qquad\qquad $\hat y_{ie}^{kr} = \sum_e S_{ie}^{kr}t_{ie}^{kr}$\\
%\qquad\qquad $\hat y_{ie}^{k} = \sum_d \hat y_{ie}^{kr}$\\
%\qquad\qquad $g_{data}^{jb} = M\sum_{kdi} a_{ij}^{kr}\left( \frac{y_i^k}{\hat y_i^{k}}-1\right) \sum_e S_{ie}^{sd} f_e^m t_{ie}^{kr}$\\
%\qquad\qquad $g_{reg}^{jb} = \sum_j \sum_{j' \in \mathcal{V}_j}\lambda^m \psi'(\rho_j^{m,(n)}-\rho_{j'}^{m,(n)})$\\
%\qquad\qquad $H_{data}^{juv} = M\sum_{kdi}  a_{ij}^{kr}A_{j}^{kr} \sum_{e}S_{ie}^{kr} f_e^u f_e^v t_{ie}^{kr}$\\
%\qquad\qquad $H_{reg}^{juv} = \sum_j^J \sum_{j' \in \mathcal{V}_j}2\lambda^m \psi''(\rho_j^{u}-\rho_{j'}^{u})$ \qquad if $u \neq v$\\
%
%
%\qquad\qquad for each voxel $j=1:J$\\
%\qquad\qquad\qquad $g_j^{m}= g_{data}^{jb}+g_{reg}^{jb}$\\
%\qquad\qquad\qquad $H_j^{m}={\rm Matrix}_{uv}\{H_{data}^{juv}+H_{reg}^{juv}$\}\\
%\qquad\qquad\qquad $\rho_j^{m}=\rho_j^{m}-{(H_j^{m})}^{-1}g_j^{m}$\\
%\\
%\\
%\\
%\\
%\\
%\qquad\qquad end \\
%\qquad end \\
%end\\
%
%
%
%\hline
%
%\end{tabular}
%\label{tab1}
%\end{table}

\subsection{Experiments}
Two experiments were performed on our in-house CBCT imaging benchtop. The X-ray source is a rotating-anode diagnostic grade X-ray tube (Varex G-242, Varex Imaging Corporation, UT, USA). In the first experiment, a SL-FPD (Varex 4343CB, Varex Imaging Corporation, UT, USA) with native pixel dimension of 0.139~mm $\times$ 0.139~mm and matrix dimension of $3072\times3072$ was used. It was $3\times3$ binned during the rotation-by-rotation dual-energy scans with 75~kVp (1.5 mm aluminium(Al) plus 0.2 mm Cu filtration) and 125~kVp (1.5 mm Al plus 1.2 mm Cu filtration) beam settings. The source to detector distance (SDD) was 1560.6~mm, and the source to rotation center distance (SOD) was 1156.3~mm. Projections were collected from one full $360^{\circ}$ rotation with an angular interval of $0.8^{\circ}$. In this experiment, a pig leg sample and the Catphan-700 phantom (CTP714 high resolution module) were scanned.

\begin{figure}[htb]
\centerline{\includegraphics[width=11cm]{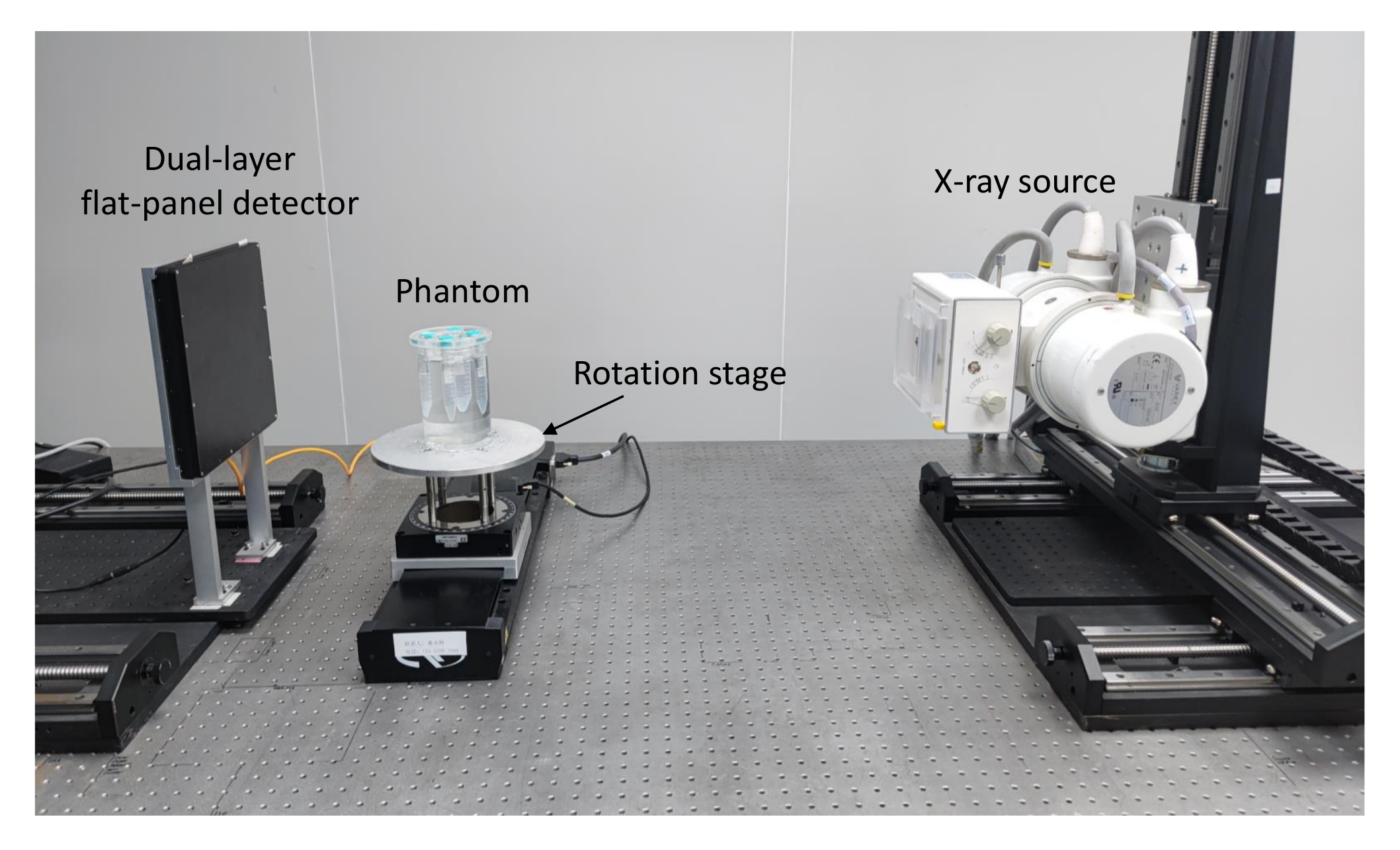}}
\caption{The experimental dual-energy CBCT imaging benchtop with a dual-layer FPD.}
\label{fig2}
\end{figure}

The second experiment was performed with a DL-FPD prototype (CareView 560RF-DE, CareRay Digital Medical Technology Co., Ltd., China), as shown in Fig~\ref{fig2}. This detector has $1536 \times 1536$ pixels with native pixel dimension of 0.154~mm $\times$ 0.154~mm. The $2 \times 2$ detector binning mode was used during the data acquisition. The source to rotation center distance was 1475.0~mm, and the source to the top and bottom detector layers were 1612.8~mm and 1619.4~mm, respectively. An additional 1.0 mm Copper filtration was inserted between the top layer and the bottom layer to harden the X-ray beam. More detailed parameter settings are listed in Table \ref{table1}. The pig leg sample and an iodine phantom were scanned. For the iodine phantom, four iodine inserts with varied concentrations (2.5 mg/cm$^3$, 5 mg/cm$^3$, 10 mg/cm$^3$, 20 mg/cm$^3$) were emerged in a cylinder water tank with diameter of 10 cm. Be aware that this DL-FPD was not worked under the newly proposed suRi method with sub-pixel shifting since such innovative data acquisition protocol is still under development by the manufacturer.

\begin{table}[htb]
\caption{Experimental data acquisition parameters.}
\label{table1}
\centering	
\begin{tabular}{p{150pt}p{160pt}p{110pt}}
\hline\hline
Detector type           & SL-FPD & DL-FPD\\ \hline
Tube voltage (kVp)       & 75; 125                                                                                              & 125                                                                    \\ 
Tube Current (mA)   & 8; 12.5                                                                                              & 12.5                                                                   \\ 
\multirow{2}{*}{Beam filtration (mm)}     & 75 kVp: 1.5 Al, 0.2 Cu & \multirow{2}{*}{0.4 Cu}\\                                                
 & 125 kVp: 1.5 Al, 1.2 Cu & \\
Total views            & 450 & 450 \\ 
SOD (mm)            & 1156.3                                                                                               & 1475.0                                                                 \\ 
\multirow{2}{*}{SDD (mm)}            & \multirow{2}{*}{1560.6} & 1612.8 (top) \\
 & & 1619.4 (bottom) \\ 
\multirow{2}{*}{Image matrix}          & Pig leg: 512$\times$512 & \multirow{2}{*}{384$\times$384}\\ 
& Catphan: 660$\times$660 &                                                                 \\ 
\multirow{2}{*}{Voxel size (mm)}     & Pig leg: 0.25$\times$0.25& \multirow{2}{*}{0.28$\times$0.28}\\ 
&Catphan: 0.31$\times$0.31 &                                                         \\ 
Detector array       & 3072$\times$3072                                                                                            & 1536$\times$1536                                                              \\ 
Pixel dimension (mm)   & 0.139$\times$0.139                                                                                                                                                                                               & 0.154$\times$0.154                                                                  \\ 
Detector binning & 3$\times$3                                                                                                  & 2$\times$2                                                                    \\ 
Manual binning      & 1$\times$2                                                                                                  & 1$\times$2                                                                    \\ 
Sub-pixel shift (mm)      & 0.417                                                                                                  & 0.308 \\ 
\multirow{2}{*}{CsI:TI thickness (mm)}    & \multirow{2}{*}{0.75} & 0.26 (top)\\ 
& & 0.55 (bottom) \\ \hline\hline
\end{tabular}
\end{table}
For both experiments, the X-ray beam was collimated into a narrow width ($<$20 mm on the detector surface) to minimize the Compton scatters. Moreover, additional $1\times2$ binning was implemented manually on the projections. To mimic the 1D sub-pixel shifting, the high-energy projection (bottom layer) is shifted by half binned pixel dimension relative to the low-energy projection (top layer) along the horizontal direction. Unless specified, the binning mode mentioned in the rest of the manuscript refers to the manual binning.

\subsection{Comparison method and parameter selection}
The conventional image domain decomposition (IDD) method was implemented for comparison. In brief, the dual-energy CT images were reconstructed with the FBP algorithm, then the density maps of two basis materials were generated via pixel-wise matrix inversion of the equation below:
	\begin{equation}
	\left[
	       \begin{array} {lr}	
			f^{1, {\rm LE}} \ f^{2, {\rm LE}} \\
			f^{1, {\rm HE}} \ f^{2, {\rm HE}} 
	       \end{array}
	\right]
	\left[
	       \begin{array} {lr}	
			\rho_1 \\
			\rho_2 
	       \end{array}
	\right]
	=
	\left[
	       \begin{array} {lr}	
			\mu^{\rm LE} \\
			\mu^{\rm HE} 
	       \end{array}
	\right],
	\label{eq29}
	\end{equation}
where $f^{m,k}$ ($m=1, 2$; $k={\rm LE, HE}$) denotes the effective mass attenuation coefficient of material $b$ at spectrum $k$, which is estimated by:
	\begin{equation}
	f^{m,k}=\frac{\sum_{e=1}^E S_e^{k} f_{e}^{m,k}}{\sum_{e=1}^E S_e^{k}}.
	\label{eq30}
	\end{equation}
Moreover, the same material decomposition algorithm used for suRi was also implemented on the aligned binning data. The obtained results are referred to as DD-align. 

Some key parameters need to be determined for the proposed suRi algorithm, including the number of sub-rays in $\mathcal{N}_i^k$, the size of neighboring pixels $\mathcal{V}_j$ in Huber regularization, the tuning parameter $\gamma$, the regularization weight $\lambda^m$ and the number of iterations $N_{\rm iter}$. Empirically, we set the size of $\mathcal{V}_j$ equal to $3\times3$. Increasing the number of sub-rays would result in more accurate imaging model, however, this would also cause intense computation burden. To balance the trade-off, 8 sub-rays was selected. The regularization weight $\lambda^m$ is critical to the image spatial resolution: the basis images would become blurred if $\lambda^m$ gets too large, and they would become very noisy if $\lambda^m$ gets too small. The specific values of $\lambda^m$, $\gamma$ and $N_{\rm iter}$ are listed in Table~\ref{table2}. Notice that the same parameters were used for the DD-align and suRi methods.

\begin{table}[htb]
\caption{Iterative parameters used by suRi.}
\label{table2}
\centering
\begin{tabular}{ccccc}
\hline\hline
\multirow{2}{*}{Parameters} & \multicolumn{2}{c}{SL-FPD CT}   & \multicolumn{2}{c}{DL-FPD CT}        \\ \cline{2-5} 
                            & Pig leg        & Catphan        & Pig leg        & Iodine              \\ \hline
$\lambda^m$                 & {[}3, 3{]}     & {[}1, 1{]}     & {[}1, 1{]}     & {[}$10^2$,$10^5${]} \\ 
$\gamma$                    & {[}0.1, 0.1{]} & {[}0.1, 0.1{]} & {[}0.1, 0.1{]} & {[}0.1, 0.001{]}    \\ 
$N_{iter}$                  & 40             & 70             & 40             & 100                 \\ \hline\hline
\end{tabular}
\end{table}

\subsection{Characterization of image spatial resolution}
To characterize the spatial resolution, the modulation transfer function (MTF) was calculated upon the reconstructed Catphan CT images. The method proposed by Droege \textit{et al.} \cite{Droege1982APM, Droege1984ModulationTF} was used to determine the MTF from the line-pair patterns:
			\begin{eqnarray}	
	{\rm MTF}(\mathcal{F}) = &\frac{\pi }{\sqrt{2}M_0}&\sqrt{\sum_{n}C_n\frac{V(n\mathcal{F})}{n^2}}, \qquad n=1,3,5,... \cr
	{\rm with} \qquad \qquad &V(n\mathcal{F})&=V'(n\mathcal{F})-V_0,\cr	
		&C_n&=\left\{
	             \begin{array} {lr}	
			-1  \quad  {\rm if} \ p' =1 \  {\rm and} \ n>1 \\
			0 \qquad {\rm if} \ p>p' \\
			1 \qquad {\rm otherwise} \\
	             \end{array}
	             \right. ,
	\label{eq31}
	\end{eqnarray}
where $\mathcal{F}$ represents the spatial frequency (unit: lp/mm), $M_0=|\mu_1-\mu_2|$ is the pixel intensity difference between the largest bar and its adjacent background regions, $V'(n\mathcal{F})$ denotes the variance measured within a region-of-interest (ROI) of the bar pattern with spatial frequency $n\mathcal{F}$, $V_0$ denotes the variance of the CT image values within an uniform ROI, $C_n$ depends on the total number of prime factors $p$ and the number of different prime factors $p'$ of the number $n$. The spatial frequency $\mathcal{F}_{20\%}$ where the MTF decreases to 0.2 was calculated.

\begin{figure*}[htb]
\centerline{\includegraphics[width=15cm]{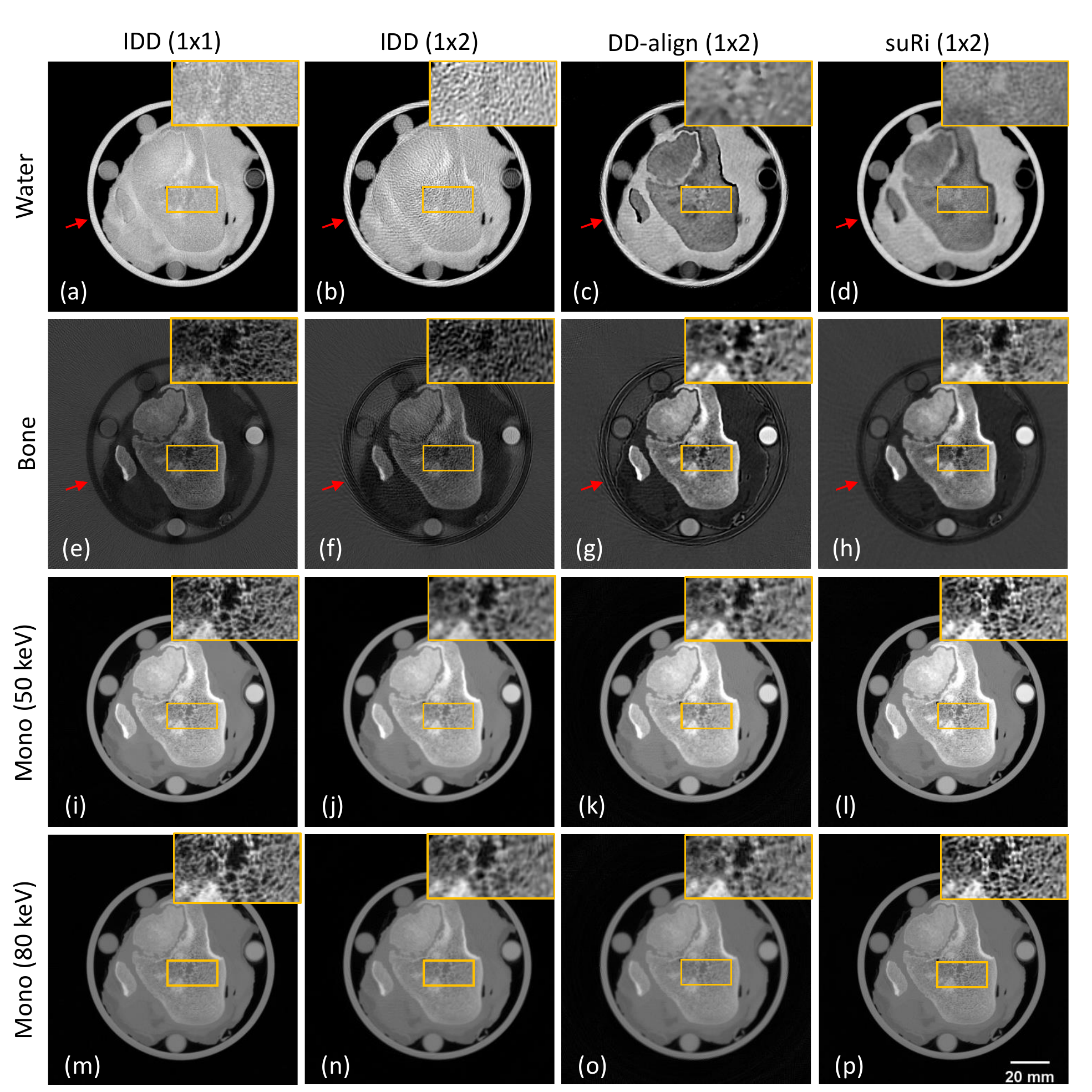}}
\caption{The decomposed dual-energy imaging results of the pig leg specimen data acquired from the SL-FPD. Images in the first column are the reference with high spatial resolution. As seen, the images in the fourth column also show very high spatial resolution as the reference images. Images in the first and second rows are the decomposed water and bone density maps, respectively. The display windows are [0.5, 1.5] g/cm$^3$ and [1, 1.5] g/cm$^3$. Images in the third and fourth rows correspond to the monochromatic images synthesized from the generated basis material images. The display window is [0, 0.45] cm$^{-1}$. The display windows for the zoomed-in ROIs in the third and fourth rows are [0.18, 0.42] cm$^{-1}$ and [0.16, 0.30] cm$^{-1}$, respectively. The scale bar denotes 20 mm.}
\label{fig3}
\end{figure*}

\section{Results} \label{sec:results}
\subsection{Results obtained from SL-FPD}
The decomposition results of the pig leg specimen are shown in Fig.~\ref{fig3}. Images in the first and second columns are reconstructed from the projections with aligned 1 $\times$ 1 binning and sub-pixel shifted 1 $\times$ 2 binning using the IDD method, images in the third and fourth columns are reconstructed from the projections with aligned 1 $\times$ 2 binning and sub-pixel shifted 1 $\times$ 2 binning using the proposed material decomposition algorithm for suRi. As seen, the water and bone basis images obtained from the IDD method contain higher image noise than those obtained from the iterative image reconstruction method. Besides, some streaking artifacts may appear in Fig~\ref{fig3}(b) and (f), as highlighted by the red arrows. These streaking artifacts are partially mitigated in the images reconstructed by the DD-align method, and most of them are efficiently suppressed by the suRi method. By comparing the monochromatic images in Fig~\ref{fig3}(i) to (p), one can clearly observe that the proposed suRi method is able to recover the fine bony structures (shown in the zoomed-in ROIs) from the 1 $\times$ 2 binned data to a similar level as those reconstructed from the 1 $\times$ 1 binned data. Without the sub-pixel shifting, however, CT images of such high spatial resolution can not be generated, see those blurry CT images obtained from the DD-align method with the aligned 1 $\times$ 2 binning data.

\begin{figure*}[htb]
\centerline{\includegraphics[width=13cm]{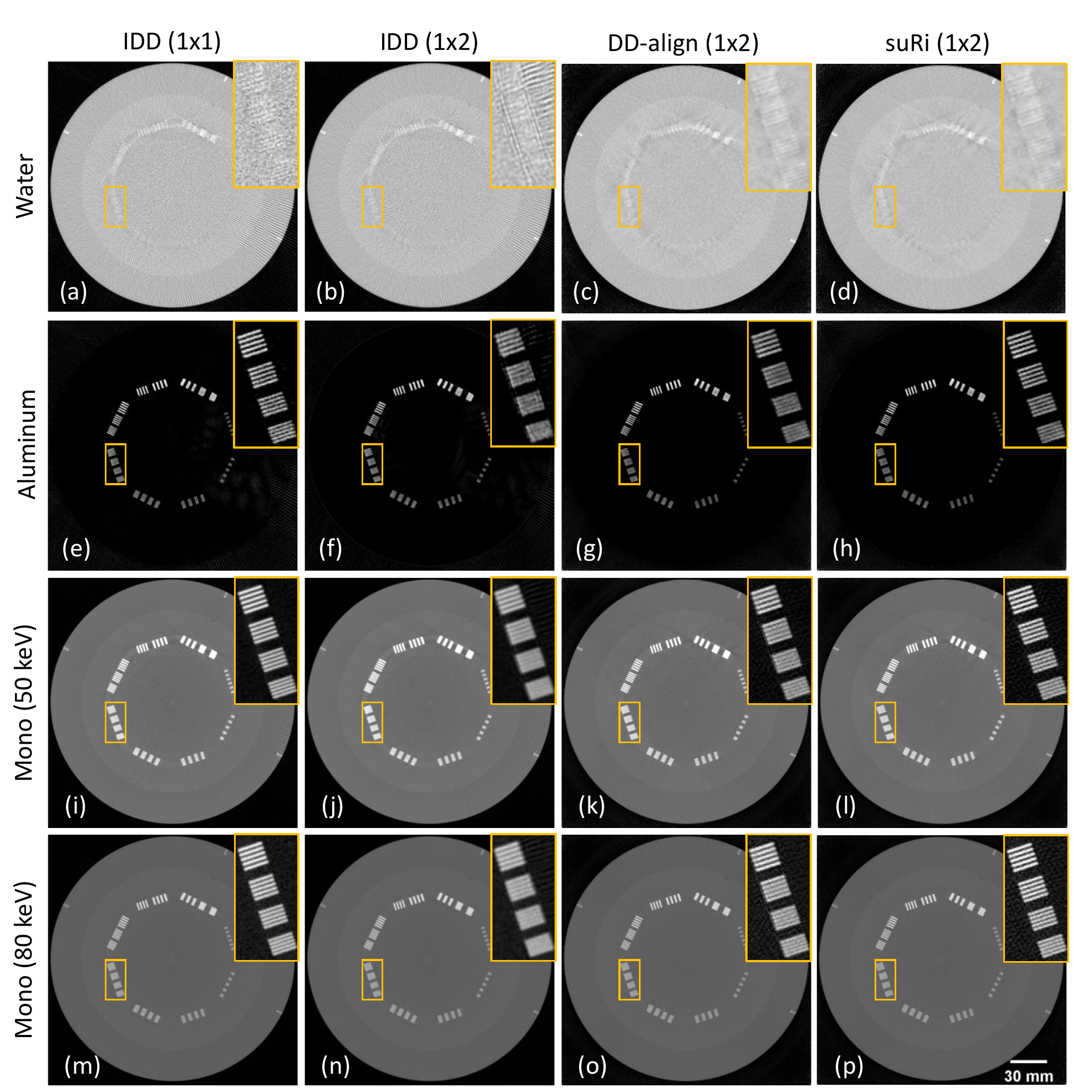}}
\caption{The dual-energy imaging results of the Catphan phantom acquired from the SL-FPD. Images in the first column are the reference with high spatial resolution. As seen, the images in the fourth column also show very high spatial resolution as the reference images. Images in the first and second rows are the decomposed water and aluminum density maps, respectively. The display windows are [0, 1.6] g/cm$^3$ and [0, 1] g/cm$^3$. Images in the third and fourth rows correspond to the monochromatic images synthesized from the generated basis images. The display window is [0, 0.5] cm$^{-1}$. The display windows for the zoomed-in ROIs in the third and last row are [0.20, 0.60] cm$^{-1}$ and [0.17, 0.39] cm$^{-1}$, respectively. The scale bar denotes 30 mm.}
\label{fig4}
\end{figure*}

\begin{figure}[htb]
\centerline{\includegraphics[width=9cm]{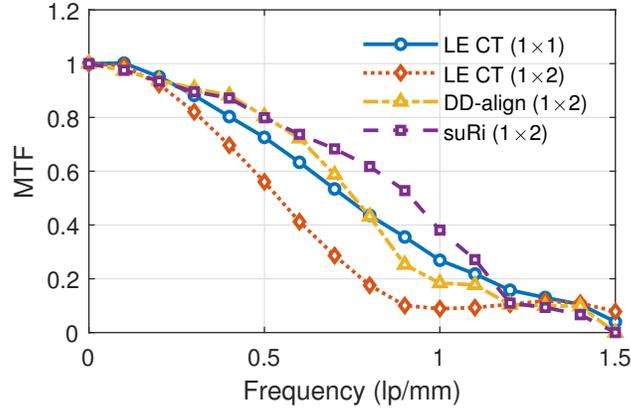}}
\caption{The MTF curves of the FBP reconstructed low-energy CT images and the monochromatic (50 keV) CT images synthesized from the DD-align and suRi method.}
\label{fig5}
\end{figure}

The imaging results of the Catphan phantom are shown in Fig.~\ref{fig4}. Similar to the pig leg results, the suRi method is capable to generate basis images and monochromatic images with higher spatial resolution than the IDD and DD-align method. The zoomed-in ROIs with line pairs in Fig.~\ref{fig4}(l) and (p) show similar visual performance compared to those in (i) and (m). The MTF curve of the dual-energy FBP images and the synthesized 50 keV monochromatic images are plotted in Fig.~\ref{fig5}. Quantitatively, the image reconstructed from the 1 $\times$ 2 binned data using the FBP method has the lowest MTF value. The penalized likelihood algorithm helps to improve the image spatial resolution: the MTF values of the DD-align and suRi method at frequency less than 0.6 lp/mm are similar, and both are higher than the FBP method (1 $\times$ 2 binned data). In regard to the high spatial frequency region, the suRi method outperforms the DD-align method due to the sub-pixel shift in the 1 $\times$ 2 binned projection data . The $\mathcal{F}_{20\%}$ of the FBP (1 $\times$ 1), FBP (1 $\times$ 2), DD-align and suRi correspond to 1.13 lp/mm, 0.78 lp/mm, 0.98 lp/mm and 1.14 lp/mm, respectively. As a results,the suRi method improves the spatial resolution by 46.15\% compared with the FBP reconstruction (1 $\times$ 2 binned) method, and 16.33\% compared with the DD-align method.

\begin{figure*}[htb]
\centerline{\includegraphics[width=18.5cm]{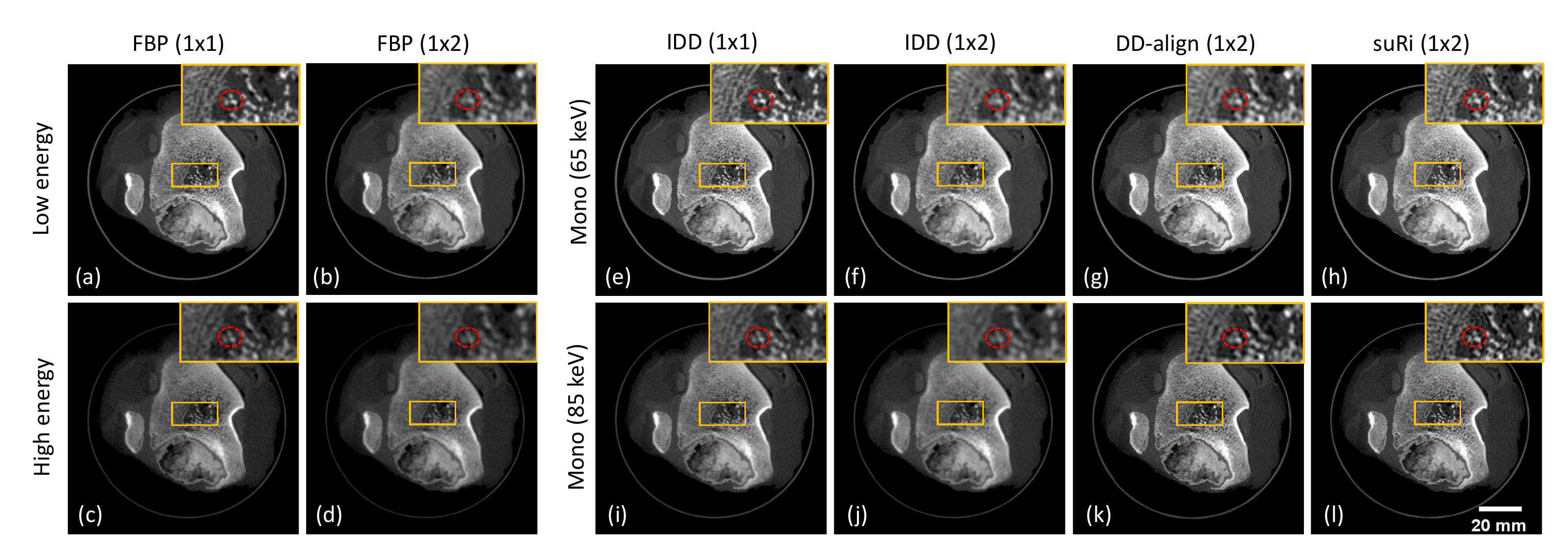}}
\caption{Imaging results of the pig leg data obtained from the DL-FPD. The FBP reconstructed low-energy and high-energy CT images are shown on the left two columns, and the synthesized monochromatic CT images using the IDD and suRi algorithms are shown on right four columns. The display window is [0.15, 0.37] cm$^{-1}$ for the low-energy CT images and the monochromatic 65 keV images, and is [0.15, 0.33] cm$^{-1}$ for the high-energy CT images and the monochromatic 85 keV images. The scale bar denotes 20 mm.}
\label{fig6}
\end{figure*}
\subsection{Results obtained from DL-FPD}
Figure~\ref{fig6} shows the dual-energy imaging results of the pig leg specimen obtained from the DL-FPD. Images in the first and second columns are dual-energy CT images reconstructed by the FBP algorithm. Images in the rest columns are monochromatic CT images synthesized from the IDD, DD-align and suRi method. Since the CsI:TI scintillator in the top and bottom layers have different thicknesses, slight spatial resolution difference in the FBP reconstructed low-energy and high-energy CT images are observed, see Fig.~\ref{fig6}(a) - (d). The synthesized monochromatic CT images from the IDD method look similar to the FBP reconstructed LE and HE CT images. Specifically, the IDD reconstructed CT images from 1 $\times$ 2 binned projections show worse spatial resolution than the results obtained from the 1 $\times$ 1 binned projections. The DD-align method (1 $\times$ 2 binned data) slightly enhances some of the fine bony structures, see the highlighted red circles, but it is still difficult to recover the full image resolution. On the contrary, the proposed suRi method generates the best CT images with the highest spatial resolution, which are visually comparable to ones reconstructed from the 1 $\times$ 1 binned projections by the FBP algorithm.

\begin{figure*}[htb]
\centerline{\includegraphics[width=13.5cm]{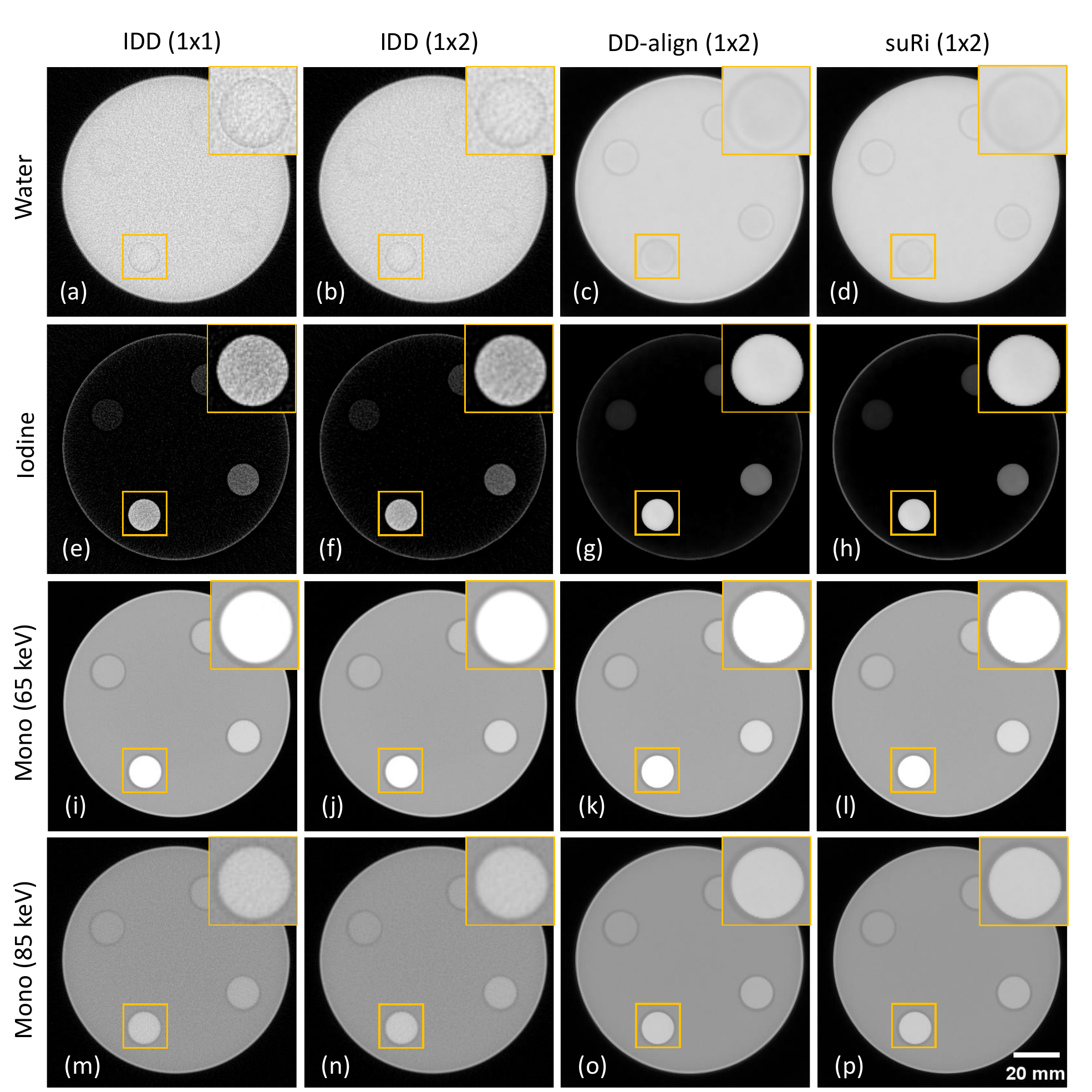}}
\caption{The dual-energy imaging results of the iodine phantom obtained from the DL-FPD. Images in the first and second rows are the decomposed water and iodine density maps, respectively. The display windows are [0, 1.2] g/cm$^3$ and [0, 25] mg/cm$^3$. Images in the third and fourth rows are the synthesized monochromatic images from the generated basis material images. The display window is [0, 0.3] cm$^{-1}$. The insert with 20 mg/cm$^3$ iodine is zoomed-in. The scale bar denotes 20 mm.}
\label{fig7}
\end{figure*}
\begin{figure}[htb]
\centerline{\includegraphics[width=9.5cm]{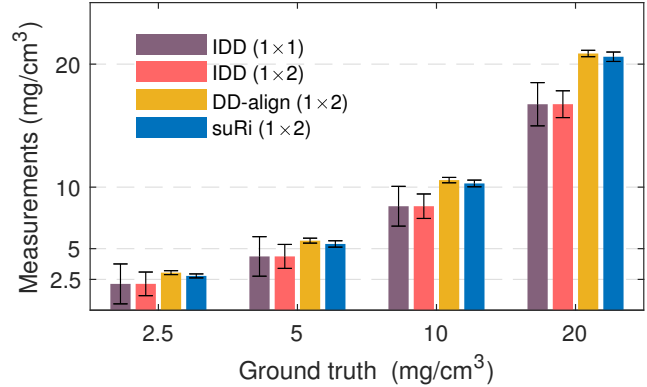}}
\caption{The mean values and standard deviations of the iodine concentrations (2.5 - 20 mg/cm$^3$) measured by the IDD, DD-align and suRi method.}
\label{fig8}
\end{figure}

The imaging results of the iodine phantom are shown in Fig.~\ref{fig7}. For all methods, the iodine material can be well separated from water. Compared with the IDD algorithm, the DD-align and suRi method could efficiently suppress the noise of the basis images. The mean values and standard deviations (STDs) of the estimated iodine concentrations are plotted in Fig.~\ref{fig8}. It is found that the IDD method slightly underestimate the iodine densities, and the decomposed results of the DD-align and suRi methods are more consistent with the ground truth. Quantitatively, the average relative errors of the four methods are -14.77\%, -14.71\%, 11.26\% and 6.26\%, respectively. The small decomposition error demonstrates the viability of the proposed reconstruction algorithm for suRi in generating accurate dual-energy basis images.

\section{Discussions and Conclusion} \label{sec:conclusion}
In this paper, a sub-pixel shifting technique, which is referred to as suRi, is firstly proposed for the DL-FPD based dual-energy CBCT to generate high spatial resolution and potentially high temporal resolution material-specific basis images. In suRi, the low-energy and high-energy projection data acquired from the top and the bottom detector layers contain varied spatial information. A dedicated penalized likelihood algorithm is developed to reconstruct high spatial resolution basis images directly from these projections containing sub-pixel shifts. Physical experiments from a SL-FPD and a DL-FPD were conducted to evaluate the performance of suRi. Results demonstrate that the proposed suRi method is capable to generate dual-energy CBCT images with high spatial resolution. Compared with the conventional FBP reconstruction (1 $\times$ 2 binning), the suRi method with sub-pixel shifted binning technique could improve the spatial resolution by $46.15\%$ (at $20\%$ MTF).

\begin{figure*}[htb]

		\begin{minipage}[b]{0.32\linewidth}
	\centerline{\includegraphics[width=5cm]{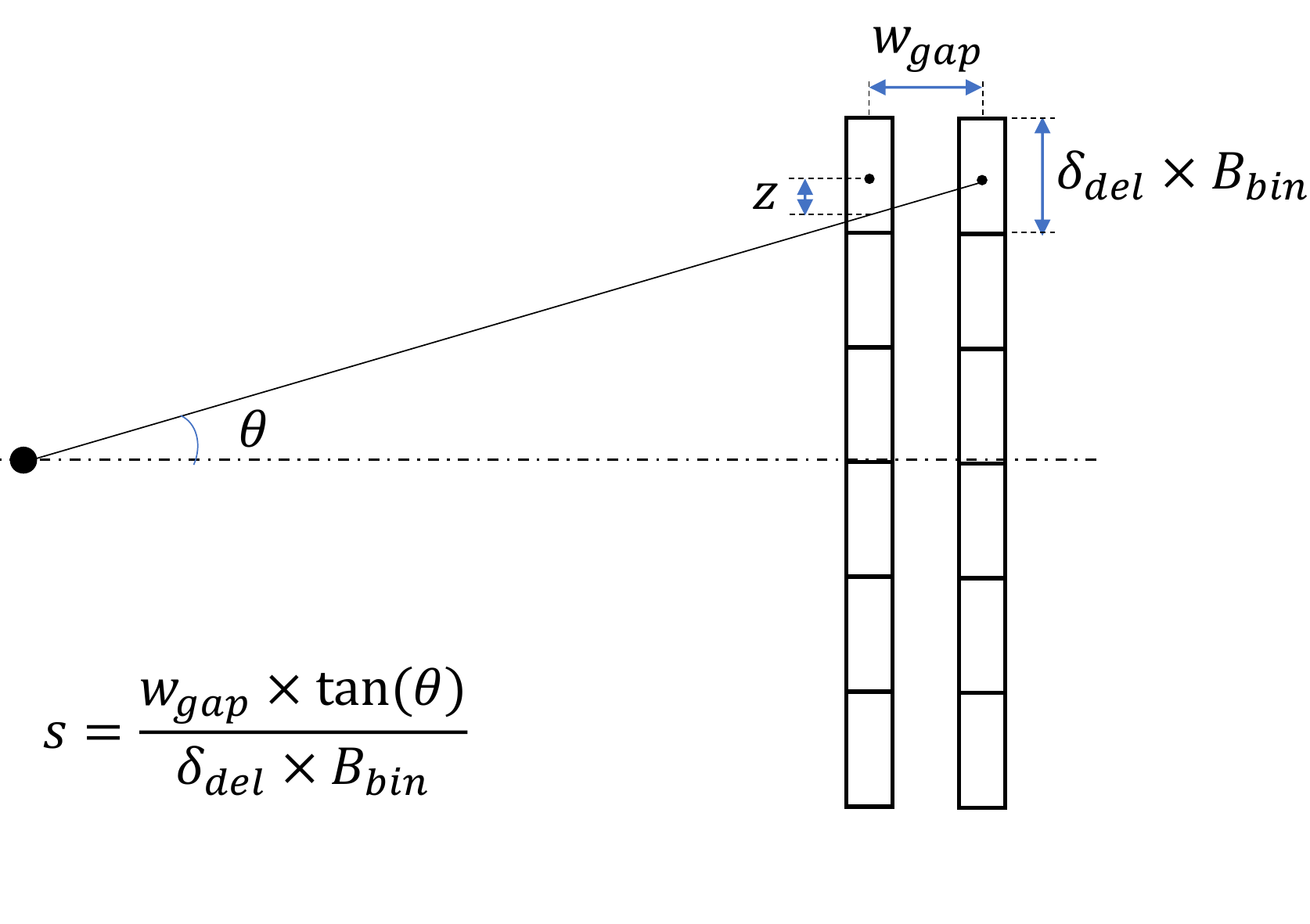}}
	 \centerline{(a)}
	\end{minipage}
	\begin{minipage}[b]{0.33\linewidth}
	\centerline{\includegraphics[width=5.5cm]{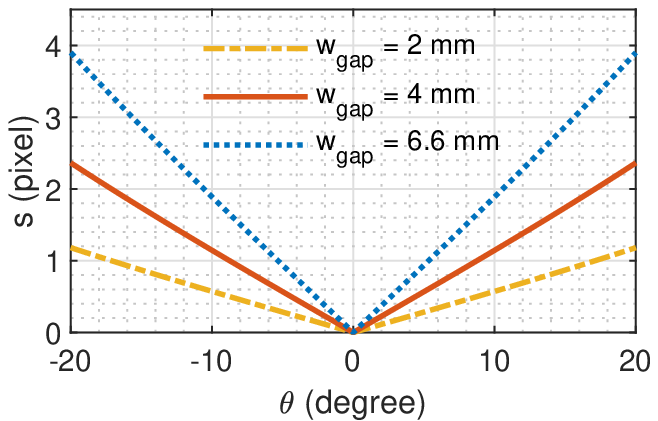}}
	 \centerline{(b)}
	\end{minipage}
	\begin{minipage}[b]{0.33\linewidth}
	\centerline{\includegraphics[width=5.5cm]{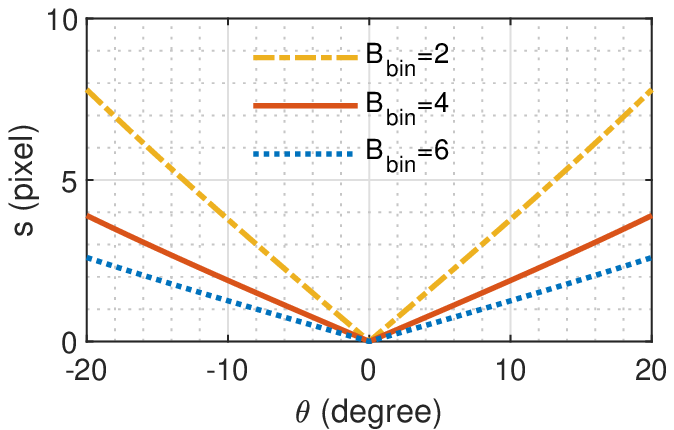}}
	 \centerline{(c)}
	\end{minipage}
\caption{(a) Illustration of the intrinsic pixel shift in dual-layer detector due to the oblique incident X-ray beam. The intrinsic pixel shift between two layers with varied (b) gap distance $w_{gap}$ and (c) pixel binning size $B_{bin}$. Notice that the DL-FPD used in this study has $w_{gap}=6.6$ mm and $B_{bin}=4$ (combination of both the detector binning and manual binning).}
\label{fig9}
\end{figure*}

The small gap (approximately 6.6 mm) between the top and bottom layers in the DL-FPD would cause intrinsic geometrical mismatches to the acquired dual-energy CBCT projections. Such geometrical mismatches correspond to the minor pixel shifts when X-rays are incident with large cone angles, see the illustration in Figure~\ref{fig9}(a). As shown in Figure~\ref{fig9}(b) and (c), such relative pixel shifts are proportional to the layer gap, but inversely proportional to the binned pixel size. As a result, the proposed suRi method would generate the highest spatial resolution improvement only in the following cases: (1) Small object size, in which the field-of-view (FOV) is mostly captured by the centeral area of the detector. (2) Small gap distance between layers. It is observed from Fig~\ref{fig9}(b) that the intrinsic pixel shift can be reduced by a factor of 3.3 if the gap distance is decreased from 6.6 mm down to 2 mm\cite{Gu2022Evaluation}. (3) Large pixel binning size. As shown in Fig~\ref{fig9}(c), this would result in reduced intrinsic pixel shift. Consequently, the suRi method would become especially useful when high acquisition speed is required.

The present work has several limitations. First, only small sized objects with 10 cm diameter were imaged by our DL-FPD to demonstrate the super resolution imaging feasibility of the proposed suRi method. This corresponds to small FOV and thus negligible intrinsic pixel shifts.
Second, only the half-pixel shift along the horizontal direction is manipulated, and the two-dimensional half-pixel shifts along both the horizontal and vertical directions have not investigated yet. It would be studied as one of the future works. 
Third, the low-energy and high-energy CT images acquired with the DL-FPD have inconsistent spatial resolutions due to the varied thickness of the scintillator in each layer. This may lead to degraded image quality of the decomposed basis materials. To overcome this problem, model-based reconstruction algorithm that considers the projection data blur\cite{Wang2021HighresolutionMM}, or novel DL-FPD structures with equal scintillator thickness\cite{Gu2022Evaluation} could be investigated.
Fourth, the proposed one-step material decomposition algorithm is sensitive to the value of the regularization weight $\lambda^m$ of the basis materials. For example, the water basis image may get blurred when $\lambda^m$ of the bone basis decreases, even if $\lambda^m$ of the water basis keeps unchanged. Careful selections of the regularization weights and development of a new cross-material regularizer\cite{Huh2011IterativeIR,Wang2020ProspectivePA} may be viable solutions. 
Fifth, we assumed the Poisson distribution of the measured data in the likelihood model. Although this is a good approximation\cite{Man2001AnIM}, the projection data may not faithfully follow the Poisson statistics because of the energy weighted signal integration in the FPD. Thus, a more realistic statistical model needs be designed in future to further improve the accuracy of the decomposed dual-energy CT images. 
Last, the potential advancements of suRi in increasing the temporal resolution for CBCT imaging is not investigated in this study. We believe this should be a very interesting topic and may bring changes to the current four-dimensional (4D) CBCT imaging applications such as the 4D neurovascular or cardiac imaging. 

In future, we believe the proposed suRi method in this work would bring many new opportunities to innovate the current CBCT imaging systems.
First, detector binning sizes that are larger than $1\times2$ (e.g., $6\times6$ or $8\times8$) could be utilized in DL-FPD to achieve even faster signal readout speed while obtaining CT images with moderate spatial resolution. If the signal readout speed is doubled, the entire CBCT scan time could be reduced by 50\% assuming the same total number of projections, or alternatively, the total number of projections could be doubled (i.e., much finer angular sampling interval) when the entire CBCT scan time is fixed. For the former case, the temporal resolution would be greatly improved \cite{Chen2015SynchronizedMR}; for the latter case, the CBCT image quality would be significantly enhanced due to the mitigation of streaking artifacts induced by the insufficient angular sampling rate \cite{Bian2010EvaluationOS}.
Second, the suRi method can also be applied on the multi-layer FPD (ML-FPD), e.g., triple-layer FPD\cite{LopezMaurino2016TheoreticalAM}, based spectral CBCT imaging system. To do that, more complex sub-pixel shifting strategy needs to be investigated and designed, as well as the reconstruction algorithm to obtain multiple sets of high resolution spatial-spectral (-temporal) CT images.
Third, the suRi method is inherently compatible with the $1\times1$ pixel binning mode in either SL-FPD or ML-FPD. For this special scenario, sub-pixel shifts among these layers have to be physically determined before laterally moving the SL-FPD \cite{Yan2015SuperRI} or assembling the DL/ML-FPD hardwares. 
Fourth, the deep learning techniques, especially the convolution neural networks (CNN), can be investigated to learn the high resolution LE and HE projections in the sinogram domain\cite{Tang2020GenerativeAN}, or to reconstruct the high resolution basis material CT images directly from the sub-pixel shifted projections\cite{Su2021DIRECTNetAU}. The utilization of CNN could significantly reduce the image reconstruction time compared with the iterative algorithms, and thus make the suRi method more viable in practice.
Fifth, the flying focal spot technique can be combined with the suRi mehtod to further improve the spatial (temporal) resolution of the reconstructed CT images.
Finally, the proposed suRi method would also benefit the DL/ML-FPDs made from the latest IGZO (indium gallium zinc oxide) based X-ray receptors.

In conclusion, a super resolution dual-energy CBCT imaging method, suRi, is developed based on the DL-FPD. It is capable to overcome the long running difficulty of achieving high spatial resolution and high signal readout speed at the same time in CBCT imaging. By using of the penalized likelihood material decomposition algorithm, experimental results demonstrate that high spatial resolution basis images can be directly reconstructed from the dual-energy projections containing sub-pixel shifts. As a result, we believe this developed suRi method would greatly enhance the medical imaging performance of the DL-FPD (or ML-FPD) based dual-energy (spectral) CBCT systems in future.

%\appendices
\setcounter{equation}{0}
\renewcommand\theequation{A.\arabic{equation}}
\section*{Appendix}\label{sec:Appendix}

The gradient of $Q_4(\boldsymbol \rho; \boldsymbol{\rho}^{(n)})$ is
	\begin{eqnarray}	
	\left.\frac{\partial Q_4(\boldsymbol \rho; \boldsymbol{\rho}^{(n)})}{\partial \rho_j^m}\right|_{\boldsymbol \rho= \boldsymbol{\rho}^{(n)}}=\cr
	\sum_{k=1}^K \sum_{r \in \mathcal{N}_i^k} \sum_{i=1}^I  a_{ij}^{kr}\left( \frac{y_i^k}{\hat y_i^{k,(n)}}-1\right)\sum_{e=1}^ES_{ie}^{kr} f_e^m t_{ie}^{kr}(\boldsymbol{\rho}^{(n)}).
	\label{eq22}
	\end{eqnarray}

The second derivative of the Hessian matrix of $Q_4(\boldsymbol \rho; \boldsymbol{\rho}^{(n)})$ is
	\begin{eqnarray}
	\left.\frac{\partial^2 Q_4(\boldsymbol \rho; \boldsymbol{\rho}^{(n)})}{\partial \rho_j^c \partial \rho_j^d}\right|_{\boldsymbol \rho= \boldsymbol{\rho}^{(n)}}=  
	\sum_{k=1}^K \sum_{r \in \mathcal{N}_i^k} \sum_{i=1}^I  a_{ij}^{kr}\left(\sum_{i=1}^J a_{ij}^{kr}\right) \sum_{e=1}^E  \frac{ S_{ie}^{kr}T_{ie}^{kr,(n)}}{ \theta_i^{kr,(n)} \beta_{ie}^{kr,(n)}}f_e^c f_e^d . 
	\label{eq23}
	\end{eqnarray}

By substituting the curvature $T_{ie}^{kr,(n)}$ with the second derivative ${\partial^2 g_{ie}^{kr}(l_{ie}^{krm})}/{\partial l_{ie}^{krc} \partial l_{ie}^{krd}}=\hat y_i^{k,(n)}$, and combining the definition of $\theta_i^{kr,(n)}$ and $\beta_{ie}^{kr,(n)}$ in \eqref{eq11}, we have:
		\begin{eqnarray}	
	\left.\frac{\partial^2 Q_4(\boldsymbol \rho; \boldsymbol{\rho}^{(n)})}{\partial \rho_j^c \partial \rho_j^d}\right|_{\boldsymbol \rho= \boldsymbol{\rho}^{(n)}}= \sum_{k=1}^K \sum_{r \in \mathcal{N}_i^k} \sum_{i=1}^I  a_{ij}^{kr}\left(\sum_{i=1}^J a_{ij}^{kr}\right) \times \cr
	 \sum_{e=1}^ES_{ie}^{kr} f_e^c f_e^d  {\rm exp} \left(-\sum_{m = 1}^M f_e^m \sum_{j=1}^J a_{ij}^{kr} \rho_j^{m,(n)}\right).
	\label{eq24}
	\end{eqnarray}
The gradient of the surrogate function of the regularization term is
		\begin{equation}	
	\left.\frac{\partial \mathcal{R}(\boldsymbol \rho; \boldsymbol{\rho}^{(n)})}{\partial \rho_j^m}\right|_{\boldsymbol \rho= \boldsymbol{\rho}^{(n)}}=\sum_j^J \sum_{j' \in \mathcal{V}_j}\lambda^m \psi'(\rho_j^{m,(n)}-\rho_{j'}^{m,(n)}),
	\label{eq25}
	\end{equation}
with
	\begin{equation}
	\psi'(\Delta)= \left\{
	             \begin{array} {lr}	
			\Delta \qquad \  {\rm if}\ |\Delta| \leq \gamma \\
			 |\gamma| \qquad {\rm if}\ |\Delta| > \gamma
	             \end{array}
	             \right. .
	\label{eq26}
	\end{equation}
The second derivative of the Hessian matrix of $\mathcal{R}(\boldsymbol \rho; \boldsymbol{\rho}^{(n)})$ is denoted as:
	\begin{equation}
	\left.\frac{\partial^2 \mathcal{R}(\boldsymbol \rho; \boldsymbol{\rho}^{(n)})}{\partial \rho_j^c \partial \rho_j^d}\right|_{\boldsymbol \rho=\boldsymbol{\rho}^{(n)}}= \left\{
	             \begin{array} {lr}	
			\sum_j^J \sum_{j' \in \mathcal{V}_j}2\lambda^m \psi''(\rho_j^{c,(n)}-\rho_{j'}^{d,(n)}) \\ \qquad \qquad \qquad \  {\rm if}\ c=d \\
			 0 \qquad \qquad \qquad {\rm if}\ c \neq d
	             \end{array}
	             \right. 
	\label{eq27}
	\end{equation}
with
	\begin{equation}
	\psi''(\Delta)= \left\{
	             \begin{array} {lr}	
			1 \qquad  {\rm if}\ |\Delta| \leq \gamma \\
			0 \qquad  {\rm if}\ |\Delta| > \gamma
	             \end{array}
	             \right. .
	\label{eq28}
	\end{equation}

\bibliography{mybib}% Produces the bibliography via BibTeX.

\end{document}